\documentclass[twocolumn,amsmath,amssymb,prb]{revtex4-2}

\usepackage{titlesec}
\usepackage[colorlinks,bookmarks=false,citecolor=blue,linkcolor=red,urlcolor=blue]{hyperref}
\usepackage{epsfig}
\usepackage{color}
\usepackage{subfigure}
\usepackage{graphicx}    
\usepackage{dcolumn}     
\usepackage[version=4]{mhchem}
\usepackage{xcolor}

\begin{document}

\title{Quantum chemical insights into hexaboride electronic structures: \\
       correlations within the boron \textit{p}-orbital subsystem}

\author{Thorben Petersen}
\email{t.petersen@ifw-dresden.de}
\author{Ulrich K.~R{\"o}{\ss}ler}
\author{Liviu Hozoi}
\email{l.hozoi@ifw-dresden.de}

\affiliation{Institute for Theoretical Solid State Physics, Leibniz IFW Dresden, 
    Helmholtzstr.~20, 01069 Dresden, Germany}

\begin{abstract}
\noindent
The notion of strong electronic correlations arose in the context of $d$-metal oxides such as NiO 
but can be exemplified on systems as simple as the H$_2$ molecule.
Here we shed light on correlation effects on B$_6^{2-}$ clusters as found in $M$B$_6$ hexaborides
and show that the B 2$p$ valence electrons are fairly correlated.
B$_6$-octahedron excitation energies computed for CaB$_6$ and YbB$_6$ agree with peak positions 
found by resonant inelastic x-ray scattering, providing a compelling picture for the latter.
Our findings characterize these materials as very peculiar $p$-electron correlated systems and 
call for more involved many-body investigations within the whole hexaboride family, 
both alkaline- and rare-earth compounds, not only for $N$- but also $(N\!\pm\!1)$-states 
defining e.\,g. band gaps.
\end{abstract}


\date\today
\maketitle

\section*{Introduction}

The physical properties of $M$B$_6$ hexaborides are remarkably diverse and have received persistent
interest \cite{inosov2021rare,Cahill2019}, as a thorough understanding of this peculiar family of
materials could not be achieved yet.
Some of its members possess properties relevant for applications, generally determined by specificities
of their electronic structures: CaB$_6$ is a semiconductor \cite{CaB6_denlinger_2002}, YB$_6$ turns
superconducting for temperatures lower than 8 K \cite{YB6_junod_2006}, LaB$_6$ is widely used as
thermionic electron emitter \cite{LaB6_swanson_1981}, EuB$_6$ displays colossal magneto-resistance
\cite{EuB6_fisk_1998}, the heavy-fermion system Ce$_{1-x}$La$_x$B$_6$ has been intensively studied in the
context of multipolar phases and magnetically hidden orders \cite{CeB6_inosov_2016,CeB6_thalmeier_2019},
while SmB$_6$
%
is believed to host correlated topological states \cite{SmB6_allen_2020}.
However, even for the seemingly simplest member of the family, CaB$_6$, there are important
electronic-structure features that are not at all clear, for example, the size and nature of its band gap: 
density-functional computations yield a vanishing band gap \cite{CaB6_hasegawa_1979,
CaB6_massidda_1997,CaB6_pickett_2000,CaB6_aryasetiawan_2002,CaB6_kino_2002}
while experiment indicates a gap of $\sim$1 eV \cite{CaB6_denlinger_2002}.
The band structures of SmB$_6$ and YbB$_6$ are also a topic of active debate, especially in relation
to their surface states \cite{SmB6_allen_2020,YbB6_hasan_2015,YbB6_chen_2020,YbB6_min_2016}; 
for SmB$_6$, even predictions made for the symmetry of the \ce{Sm^3+}-ion ground-state term 
are recently contradicted by experiment \cite{SmB6_tjeng_2018,SmB6_tjeng_2019}.

In this context, we argue that one aspect constantly neglected in the study of these compounds is
correlations within the B 2$p$ electronic subsystem.
To address such physics, we here employ \textit{ab initio} multi-configurational wavefunction theory.
For CaB$_6$ and YbB$_6$, representative divalent hexaborides, the ground-state wavefunctions turn
out strongly multi-configurational, with large admixture of higher-lying configurations.
Trying to understand the nature of B 2$p$ $N$-particle states, a rich $N$-particle excitation spectrum
is evidenced in the \textit{ab initio} calculations.
These computed $p$-$p$ excitation energies compare well with peak positions in B $K$-edge
RIXS spectra reported by Denlinger \textit{et al.}\;\cite{CaB6_denlinger_2002,YbB6_denlinger_2002}.
%
Analysis of the many-body wavefunctions in terms of site-centered valence orbitals suggests that the 
\textit{ab initio} data can be mapped onto a six-site, half-filled effective Hubbard model.
Unfolded to the extended lattice, this might provide a convenient frame for further insights into the
electronic properties of the B 2$p$ subsystem.
The computationally more expensive option is using quantum chemical methods to determine the nature
of $(N\!\pm\!1)$ quasi-particle states \cite{CuO_hozoi_08,MgO_hozoi_07,diamond_alex_14} defining e.\,g.
band gaps.

\section*{Results}

\subsection*{Correlated electronic structure of CaB$_6$}

\begin{figure}[!b]
  \centering
  \includegraphics[width=0.96\columnwidth]{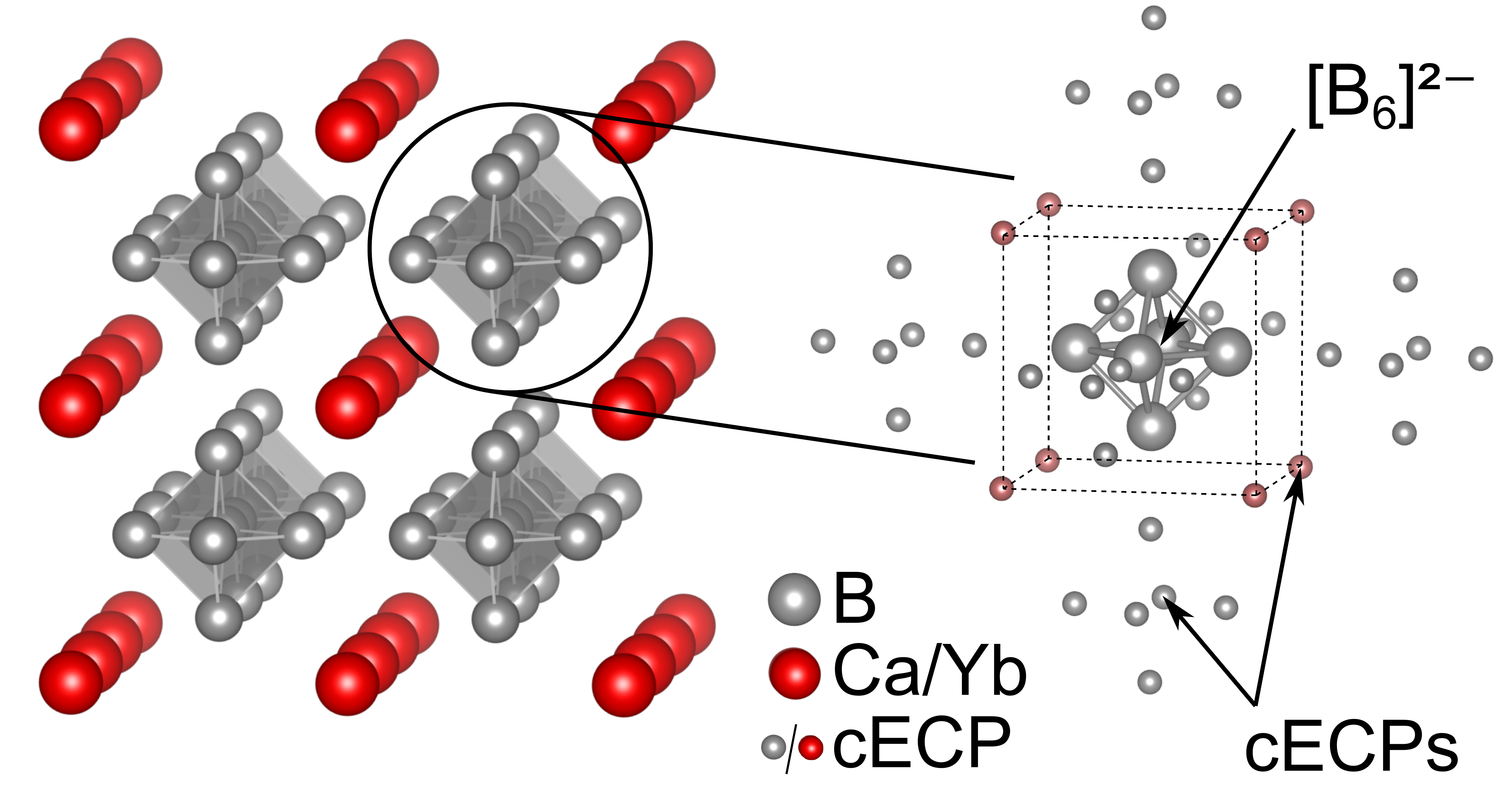}
  \caption{
    \textbf{Cubic crystal structure of $\mathbf{M}$B$_6$ hexaborides and B$_6^{2-}$ quantum mechanical cluster.}
    The quantum cluster is surrounded by 44 nearby ions modeled as capped effective core potentials
    (cECPs).
    Dashed lines highlight the cubic $M_8$ cage around a given B$_6$ octahedron 
    (red: Ca/Yb, light grey: B).
    }
  \label{fig:crystal}
\end{figure}

The crystal structure of cubic $M$B$_6$ hexaborides is CsCl-like, with $Pm{\bar{3}}m$ 
symmetry \cite{inosov2021rare,Cahill2019}.
A given B$_6$ octahedron is surrounded by eight $M$ nearest-neighbor sites defining a cube (see Fig.\;\ref{fig:crystal}).
$M$ can be an alkaline-earth atom, Y, or a rare-earth ion (e.g., La, Ce, Sm, Yb). 
Interestingly, even for CaB$_6$, a conventional closed-shell $p$-electron compound at first sight, 
standard band-structure calculations and experiment provide conflicting results on the fundamental gap: 
computations within the local density approximation (LDA) of density functional theory, for instance, 
indicate a metallic ground state \cite{CaB6_hasegawa_1979,CaB6_massidda_1997,CaB6_pickett_2000,CaB6_aryasetiawan_2002,CaB6_kino_2002} 
while a gap of $\sim$1~eV is found experimentally \cite{CaB6_denlinger_2002}.
One obvious question is then what kind of correlations are missed in (LDA-based) density functional
theory in this class of materials but the answer is not evident.
From post-LDA $GW$ calculations, both finite \cite{CaB6_bobbert_2001} and
vanishing \cite{CaB6_aryasetiawan_2002,CaB6_kino_2002} band gaps were reported.
These contradictory sets of data suggest that the results are sensitive to specific details in the 
numerical implementation.
Moreover, the $GW$ scheme is not designed for dealing with short-range correlations; the latter turn
out to be massive, as discussed below. 

For an isolated B atom, the Aufbau principle implies an $[1s^2]2s^22p^1$ valence configuration;
on a B$_6^{2-}$ octahedral unit, there are $6\!\times\!3+2\!=\!20$ electrons associated with the B
2$s$ and 2$p$ shells.
But those 20 electrons can be now regarded as occupying symmetry-adapted orbitals distributed over
several B sites of a given octahedron.
In cubic geometry (i.\,e., regular octahedron), there are $a_\text{1g}$, $e_\text{g}$, and $t_\text{1u}$
symmetry-adapted functions associated with the six $s$ atomic orbitals and $a_\text{1g}$, $e_\text{g}$,
$t_\text{1u}$, and $t_\text{2g}$ linear combinations related to the set of 18 $p$ atomic orbitals. 

To gain insight into the B 2$s$/2$p$ electronic structure, we here rely on \textit{ab initio}
quantum chemical methods \cite{Helgaker2012} as implemented in 
the \textsc{Orca} program package v5.0 \cite{Neese2020}.
The material model is a point-charge embedded \ce{B6^2-} cluster, for both \ce{CaB6} and \ce{YbB6}
(see Fig.\:\ref{fig:crystal}).
Details on the technicalities of this embedding process are given in the Methods section.
The quantum chemical investigation was initiated as a preliminary Hartree-Fock (HF) calculation that
provides starting orbitals for a much more sophisticated post-HF treatment: 
complete-active-space self-consistent-field (CASSCF) computations and subsequent multi-reference
configuration interaction (MRCI) with single and double excitations \cite{Helgaker2012}. 
To make the highly-demanding MRCI calculations feasible, we enabled the use of symmetry, according
to the $D_{2h}$ subgroup; B $1s$ orbitals were not correlated.

\begin{figure}[!t]
    \centering
    \includegraphics[width=0.95\columnwidth]{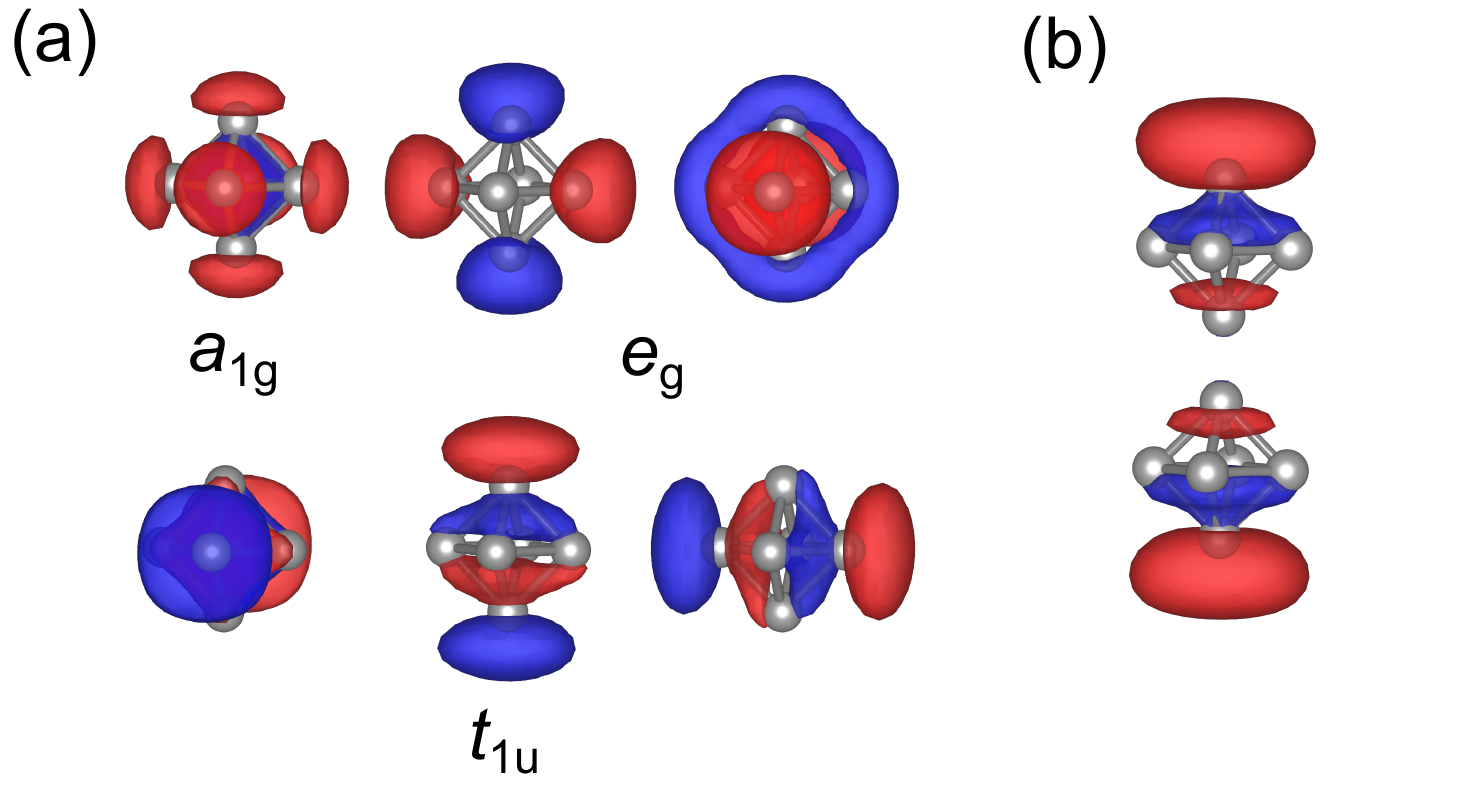}
    \caption{
     \textbf{Orbitals of the chosen complete active space with six electrons and six orbitals (CAS(6,6)).}
     (a) Symmetry-adapted ($O_h$ point-group symmetry) orbitals for the B$_6^{2-}$ unit.
     (b) Site-centered symmetry-equivalent orbitals derived from the former (only two out of six are shown).
    }
    \label{fig:act-orbs}
\end{figure}

Starting from the closed-shell HF solution, basic information on the nature of the low-lying excited
states and on the nature of most important correlation effects was gathered from a series of CASSCF
calculations in which different numbers of orbitals were considered as active.
In quantum chemical terminology, ``active'' is the set of orbitals within which all possible determinants
are generated on the basis of a given number of electrons.
These are chemically or physically relevant orbitals, to e.g.~bonding or spectroscopy, respectively;
less relevant orbitals are assigned either double (core/semi-core levels) or zero (virtual levels)
occupancy. 
By analyzing different active orbital spaces, the six molecular orbitals shown in Fig.\;\ref{fig:act-orbs}(a)
were found to be strongly correlated;
in $O_h$ point group symmetry, relevant to a B$_6^{2-}$ unit in cubic hexaborides, these orbitals
transform according to the $A_\text{1g}$, $E_\text{g}$, and $T_\text{1u}$ irreducible representations.
We found that for the aspects discussed here a six-orbital active space ($a_\text{1g}$\,+\,$e_\text{g}$\,+\,$t_\text{1u}$) accommodating six electrons
(for lower-energy orbitals on the \ce{B_6^2-} octahedron, double occupancy is imposed in CASSCF)
already provides a balanced description; we refer to this active space as CAS(6,6).

Using this active space, a total of one septet, five quintets, 15 triplets and eight singlet states 
were considered in a state-averaged \cite{Helgaker2012} CASSCF calculation.
Results computed this way for \ce{CaB6} are listed in Table\;\ref{tab:cab6}.
Electronic configurations are here expressed in terms of symmetry-adapted molecular-like orbitals
($a_\text{1g}$, $e_\text{g}$, $t_\text{1u}$);
this is common practice when discussing the electronic structure of octahedral atomic units and
also the usual way output numerical data are printed by quantum chemical programs.
A first observation is the pronounced multi-configurational structure of the computed wavefunctions.
The leading ground-state configuration, for example, has a weight of only 61$\%$ in the many-body
ground-state wavefunction ($^1\!A_\text{1g}$ state) and a similar situation is observed for many of
the excited states; the remaining weight has to do in each case with other electron configurations.
The  multi-configurational character of most of the states in Table\;\ref{tab:cab6} is also 
reflected in the natural-orbital occupation numbers, obtained through diagonalization of 
the density matrix \cite{Helgaker2012}: for the $^1\!A_\text{1g}$ ground state, for instance, 
the $a_\text{1g}$, $e_\text{g}$, and $t_\text{1u}$ occupation numbers are 1.85, 1.70 ($\times$2),
and 0.25 ($\times$3), respectively;
at the HF level the occupation of the antibonding (see Fig.\;\ref{fig:act-orbs}(a)\,) $t_\text{1u}$
levels is 0 since the HF ground-state wavefunction is 100\% $a_\text{1g}^2\:e_\text{g}^4$.
Also illustrative for the effect of electronic correlations:
the (restricted open-shell) HF\ $^1\!A_\text{1g}$--\,$^5T_\text{2g}$ splitting, for instance, is less
than 1 eV, much smaller than the CASSCF/MRCI values provided in Table\;\ref{tab:cab6}.

\begin{table}[!t]
    \caption{
        Excitation energies (eV) for a B$_6^{2-}$ octahedral block in \ce{CaB6} (CAS(6,6) reference).
        Weights of leading configurations are provided within brackets.
        Computational results based on complete active space self-consistent field (CASSCF) 
        and more sophisticated multi-reference configuration interaction (MRCI) are given 
        in comparison to experimental resonant inelastic x-ray scattering (RIXS) peaks.
    }
    \begin{ruledtabular}
        \begin{tabular}{l l c c c }
            State & Leading config. & $E_\text{CASSCF}$ & $E_\text{MRCI}$ & $E_\text{RIXS}$ \cite{CaB6_denlinger_2002} \\
            \colrule \\[-0.3cm]
            $^1A_\text{1g}$ & $a_\text{1g}^2 \: e_\text{g}^4 \: t_\text{1u}^0$ & 0 (61\%)    & 0 (61\%)    & 0   \\[0.2cm]
            $^3T_\text{2u}$ & $a_\text{1g}^2 \: e_\text{g}^3 \: t_\text{1u}^1$ & 0.99 (71\%) & 1.09 (68\%) & 1.15\\[0.2cm]
            $^5T_\text{2g}$ & $a_\text{1g}^2 \: e_\text{g}^2 \: t_\text{1u}^2$ & 2.11 (79\%) & 2.28 (75\%) & 2.6\\[0.1cm]
            $^3T_\text{2g}$ & $a_\text{1g}^2 \: e_\text{g}^2 \: t_\text{1u}^2$ & 2.24 (76\%) & 2.41 (71\%) & \\[0.1cm]
            $^1T_\text{2g}$ & $a_\text{1g}^2 \: e_\text{g}^2 \: t_\text{1u}^2$ & 2.30 (75\%) & 2.49 (69\%) & \\[0.1cm]
            $^3T_\text{1u}$ & $a_\text{1g}^2 \: e_\text{g}^1 \: t_\text{1u}^3$ & 3.90 (86\%) & 2.40 (74\%) & \\[0.1cm]
            $^1T_\text{2u}$ & $a_\text{1g}^2 \: e_\text{g}^3 \: t_\text{1u}^1$ & 4.02 (87\%) & 2.48 (75\%) & \\[0.2cm]
            $^3T_\text{1g}$ & $a_\text{1g}^2 \: e_\text{g}^2 \: t_\text{1u}^2$ & 4.66 (76\%) & 3.42 (77\%) & 3.6 \\[0.1cm] 
            $^1A_\text{1u}$ & $a_\text{1g}^2 \: e_\text{g}^1 \: t_\text{1u}^3$ & 3.72 (74\%) & 4.06 (68\%) & \\[0.1cm]
            $^3E_\text{u}$  & $a_\text{1g}^2 \: e_\text{g}^1 \: t_\text{1u}^3$ & 3.76 (70\%) & 4.11 (66\%) & \\[0.1cm] 
            $^3A_\text{1u}$ & $a_\text{1g}^2 \: e_\text{g}^1 \: t_\text{1u}^3$ & 3.77 (70\%) & 4.12 (66\%) & \\[0.1cm] 
            $^5E_\text{u}$  & $a_\text{1g}^2 \: e_\text{g}^1 \: t_\text{1u}^3$ & 3.88 (61\%) & 4.24 (61\%) & \\[0.2cm]
            $^7A_\text{1u}$ & $a_\text{1g}^1 \: e_\text{g}^2 \: t_\text{1u}^3$ & 4.53 (100\%) & 5.46 (83\%) & \\
        \end{tabular}
    \end{ruledtabular}
    \label{tab:cab6}
\end{table}

An instructive exercise is re-expressing the CAS(6,6) ground-state wavefunction in terms of 
site-centered orbitals.
To achieve this, the symmetry-adapted $a_\text{1g}$, $e_\text{g}$, and $t_\text{1u}$ composites were
transformed to a set of six symmetry-equivalent functions using the orbital ``localization'' module
in \textsc{Orca} \cite{Neese2020}.
Two of these symmetry-equivalent site-centered orbitals are displayed in Fig.\;\ref{fig:act-orbs}(b).
Their asymmetric shape clearly indicates $p$-$s$ mixing: dominant $p$ but also $s$ character, with
substantial weight at the opposite B site.
In this basis, the ground-state wavefunction amounts to 
54\% {$\lvert\uparrow\downarrow\uparrow\downarrow\uparrow\downarrow\rangle\!$},
i.\,e., 54\% determinants with single orbital occupation, and
36\%\,  $\lvert\uparrow\downarrow\uparrow\downarrow\!0\,2\rangle\!+...$\,,
 7\%\,  $\lvert\uparrow\downarrow\!0\,2\,0\,2\rangle\!+...$\,, and
 3\%\,  $\lvert0\,2\,0\,2\,0\,2\rangle\!+...$\, configurations,
where $\uparrow$, $\downarrow$, 2, and 0 stand for spin-up electron, spin-down electron, zero orbital
occupation, and double occupation, respectively.
As such, the wavefunction can in principle be mapped onto an effective six-site Hubbard-type model ($t$-$U$
or $t$-$U$-$V$).
Estimating  hopping ($t$) and Coulomb repulsion ($U$,\;$V$) integrals \cite{NiO_wim_88,CuO_martin_93,
CuO_niko_18} defining such effective models is beyond the scope of this letter.
But the quantum chemical wavefunctions and excitation energies reported here provide useful \textit{ab initio}
benchmarks for tuning the relevant parameters.
Remarkably, $p$-electron Hubbard-$U$ values as large as 9\:eV were estimated for e.\,g.~carbon and
phosphorus allotropes \cite{Wehling2011,P_3p_luis_2017}.

Good insight into correlation effects on the B$_6^{2-}$ unit is attained by inspecting the wavefunction
describing two opposite B sites (the hopping integral is larger for opposite-corner $sp$ hybrids).
To this end the following numerical test can be made:
allow all possible occupations for two $sp$ hybrids on opposite corners of the octahedron (c.f. those
pictured in Fig.\;\ref{fig:act-orbs}(b)), but restrict the occupation of each of the other four $sp$
hybrids to 1 (without re-optimizing the orbitals).
The weight of configurations with 0,\,2 and 2,\,0 orbital occupation for the opposite-corner orbitals
is only 19\%, indicative of sizable correlations.\
For comparison, in the H$_2$ molecule at equilibrium H-H interatomic distance the weight of those ionic
configurations in the HF wavefunction (uncorrelated limit) is 50\%, being reduced to about 44\% in the
exact wavefunction \cite{Jensen2007}.

\begin{figure*}[!t]
  \centering
  \includegraphics[width=0.85\textwidth]{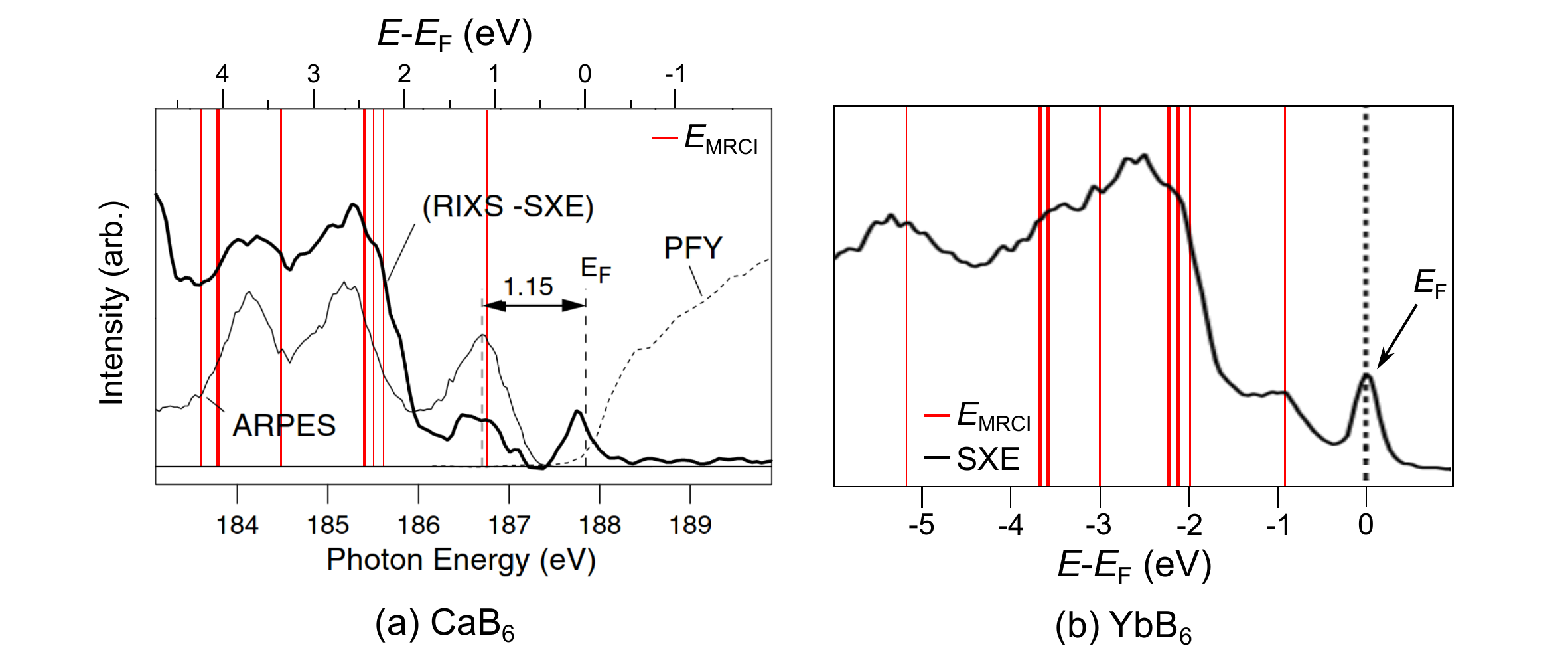}
  \caption{
    \textbf{Comparison of experimental X-ray spectra and calculated MRCI relative energies.}
    A series of experiments was conducted for (a) CaB$_6$ and (b) YbB$_6$, 
    involving angle-resolved photoemission spectroscopy (ARPES), surface x-ray photoemission (SXE),
    resonant-inelastic x-ray scattering (RIXS), and partial fluorescence-yield (PFY) spectroscopy 
    (in black). 
    Multi-reference configuration interaction (MRCI) energies are given for comparison (in red) 
    and all energies are referenced onto the Fermi energy $E_\text{F}$. 
    Figure 3(a) reprinted with permission from Denlinger \textit{et al.} \cite{CaB6_denlinger_2002},
    copyright (2022) by the American Physical Society, 
    and Fig. 3(b) reprinted with permission from Denlinger \textit{et al.} \cite{YbB6_denlinger_2002},
    copyright (2022) by World Scientific Publishing Co. PTE. LTD. 
    through the Copyright Clearance Center (CCC).
  }
  \label{fig:sxe}
\end{figure*}

Support to our computational results is provided by the good agreement between calculated
$N$-particle 
excitation energies and excitation energies extracted from B $K$-edge RIXS 
experiments \cite{CaB6_denlinger_2002}.
This is apparent by comparing numerical (MRCI level of theory) and experimental values in the last
two columns of Table\;\ref{tab:cab6} and Fig.\:\ref{fig:sxe}(a).
In other words, the numerically obtained excitation energies fit the measured excited-state relative
energies; this good agreement allows for an one-to-one assignment of the main features for less than
$\approx$4 eV energy loss in the RIXS spectra.
But the interpretation of the electronic ground state itself and of the lower $N$-particle charge
excitations resulting from this assignment is peculiar, with no correspondence in the extensive
literature available on $M$B$_6$ hexaborides.

That computations beyond the CAS(6,6) level are necessary is visible from the large corrections brought
through the MRCI treatment to some of the excitation energies, in particular, to the relative energies
of singlet and triplet states at 2.4--3.5 eV and of the $^7A_\text{1u}$ septet.
These corrections amount to up to 1.5 eV, see Table\;\ref{tab:cab6}, and point to configuration-interaction
(CI) effects that are not yet captured with a six-orbital active-space CASSCF. 
This is evident, e.\,g., from the substantial reduction of the weight of the leading configuration in
the CI wavefunction of each of those states.
Moreover, we found that more approximate post-CASSCF CI schemes such as spectroscopy-oriented CI
(SORCI) or second-order perturbation theory (PT2) treatments such as $N$-electron valence PT2
correction (NEVPT2) yield inaccurate excitation energies (see Supplementary Table 1).
Relative shifts of this magnitude when improving the description of electronic correlations (by
appropriately enlarging the active orbital space in CASSCF and/or post-CASSCF correlation treatment) 
are typically found in $d$-electron oxides, e.g., for $d^6$ states in LaCoO$_3$ \cite{lacoo3_hozoi_09}.

\subsection*{Correlated electronic structure of YbB$_6$}
For comparison, similar quantum chemical computations were carried out for the $N$-particle B$_6^{2-}$
electronic structure in the rare-earth hexaboride YbB$_6$.
The relative energies of the first twelve excited states and leading electronic configurations are
listed in Table\;\ref{tab:ybb6}.
In agreement with the computational results obtained for \ce{CaB6}, the ground-state wavefunction is
of strong multi-configurational character, with its leading configuration amounting to only 60\%.
Also in this case, the calculated MRCI energies can be compared with B $K$-edge X-ray spectra \cite{YbB6_denlinger_2002};
MRCI $N$-particle excitation energies are overlaid for this purpose onto the experimental plots of
Denlinger {\it et al.} \cite{YbB6_denlinger_2002} in Fig.\:\ref{fig:sxe}.
It is seen that the relative energy of the well-defined first peak is rather well reproduced in both
compounds.
At higher energies, the level structure is denser; qualitatively, there is reasonably good correspondence
between the peak found experimentally in each of the two materials at 2-3 eV and the excitations 
computed in that energy range.

\begin{table}[!t]
    \caption{
        Excitation energies (eV) computed for a B$_6^{2-}$ octahedral block in \ce{YbB6} (CAS(6,6) reference).
        Weights of leading configurations are provided within brackets.
        Computational results based on complete active space self-consistent field (CASSCF) 
        and more sophisticated multi-reference configuration interaction (MRCI) are given 
        in comparison to experimental resonant inelastic x-ray scattering (RIXS) peaks.
    }
    \begin{ruledtabular}
        \begin{tabular}{ l l c c c }
            State & Leading config. & $E_\text{CASSCF}$ & $E_\text{MRCI}$ & $E_\text{RIXS}$ \cite{YbB6_denlinger_2002} \\
            \colrule \\[-0.3cm]
            $^1A_\text{1g}$ & $a_\text{1g}^2 \: e_\text{g}^4 \: t_\text{1u}^0$ & 0 (60\%)    & 0    (60\%) & 0 \\[0.2cm]
            $^3T_\text{2u}$ & $a_\text{1g}^2 \: e_\text{g}^3 \: t_\text{1u}^1$ & 0.84 (72\%) & 0.95 (69\%) & 1.0 \\[0.2cm]
            $^5T_\text{2g}$ & $a_\text{1g}^2 \: e_\text{g}^2 \: t_\text{1u}^2$ & 1.78 (83\%) & 1.98 (77\%) & 2.5 \\[0.1cm]
            $^3T_\text{2g}$ & $a_\text{1g}^2 \: e_\text{g}^2 \: t_\text{1u}^2$ & 1.94 (79\%) & 2.14 (73\%) & \\[0.1cm]
            $^1T_\text{2g}$ & $a_\text{1g}^2 \: e_\text{g}^2 \: t_\text{1u}^2$ & 2.01 (76\%) & 2.22 (70\%) & \\[0.1cm]
            $^3T_\text{1u}$ & $a_\text{1g}^2 \: e_\text{g}^1 \: t_\text{1u}^3$ & 3.50 (84\%) & 2.15 (74\%) & \\[0.1cm]
            $^1T_\text{2u}$ & $a_\text{1g}^2 \: e_\text{g}^3 \: t_\text{1u}^1$ & 3.63 (87\%) & 2.23 (75\%) & \\[0.2cm]
            $^3T_\text{1g}$ & $a_\text{1g}^2 \: e_\text{g}^2 \: t_\text{1u}^2$ & 4.12 (79\%) & 3.03 (78\%) & 3.4 \\[0.2cm]
            $^1A_\text{1u}$ & $a_\text{1g}^2 \: e_\text{g}^1 \: t_\text{1u}^3$ & 3.28 (78\%) & 3.66 (70\%) & \\[0.1cm]
            $^3E_\text{u}$  & $a_\text{1g}^2 \: e_\text{g}^1 \: t_\text{1u}^3$ & 3.31 (74\%) & 3.70 (69\%) & \\[0.1cm]
            $^3A_\text{1u}$ & $a_\text{1g}^2 \: e_\text{g}^1 \: t_\text{1u}^3$ & 3.33 (74\%) & 3.72 (68\%) & \\[0.1cm]
            $^5E_\text{u}$  & $a_\text{1g}^2 \: e_\text{g}^1 \: t_\text{1u}^3$ & 3.42 (66\%) & 3.77 (66\%) & \\[0.2cm]
            $^7A_\text{1u}$ & $a_\text{1g}^1 \: e_\text{g}^2 \: t_\text{1u}^3$ & 4.20 (100\%) & 5.15 (83\%) & \\
        \end{tabular}
    \end{ruledtabular}
    \label{tab:ybb6}
\end{table}

The B $p$-$p$ excitation energies computed for \ce{YbB6} are slightly lower as compared to \ce{CaB6}.
The likely explanation is that 
the larger effective charge of Yb ions in \ce{YbB6} 
(2 or slightly larger than 2 according to experimental estimates
\cite{YbB6_nanba_1993,Gavilano2003,Zhou2015,YbB6_min_2016},
a representation that is also adopted for the surroundings of the B$_6$ octahedron 
in our calculations) stabilizes the $t_{1u}$ B-cluster orbitals as compared to the $e_g$ 
(also $a_{1g}$) components and reduces this way the $e_g\!\rightarrow\!t_{1u}$ 
excitation energies --- the $t_{1u}$ composites have stronger antibonding character, 
are more extended, and therefore sense more effectively the nearest-neighbor cations.


\section*{Conclusions}
While strong correlation effects within a whole manifold of electron configurations were 
already pointed out for the case of $d$-ion clusters \cite{nb4_majumdar_2004,fe4s4_sharma_14,
fe4s4_presti_19,V148_hozoi_20,Petersen2022}, we here describe similar physics occurring on $p$-electron 
octahedral clusters as found in $M$B$_6$ hexaborides.
In particular, we highlight the role of correlation effects on B$_6$ octahedral units.
For divalent CaB$_6$ 
and YbB$_6$,
%
our quantum-chemically computed valence-shell excitation energies agree well with peak positions 
in earlier B $K$-edge RIXS measurements \cite{CaB6_denlinger_2002,YbB6_denlinger_2002}, 
providing an interpretation of the RIXS spectrum.
%
Shifts to lower energies of the $N$-particle excited states in YbB$_6$ are ascribed to a relative
stabilization of the antibonding $t_{1u}$ cluster orbitals as a result of larger cation charges in
the Yb compound.
%
Given\, (i)\;the limitations of density functional theory in describing the electronic structure of
hexaboride compounds, documented in the case of CaB$_6$ in literature extending over a few decades
\cite{CaB6_denlinger_2002,CaB6_hasegawa_1979,CaB6_massidda_1997,CaB6_pickett_2000,CaB6_bobbert_2001,
CaB6_aryasetiawan_2002}, and\, (ii)\;more recent controversies on the electronic structures 
of e.\,g.~SmB$_6$ \cite{SmB6_allen_2020,SmB6_tjeng_2018,SmB6_tjeng_2019} and YbB$_6$
\cite{YbB6_hasan_2015,YbB6_chen_2020,YbB6_min_2016}, an extension of the present investigation 
of $N$-particle excitations to quasi-particle band structures is highly desirable.
The foundations of quasi-particle band-structure calculations with \textit{ab initio}
wavefunction-based methods were laid \cite{wfs_fulde_2019,MgO_hozoi_07,diamond_alex_14} 
and such an approach provides better control over the underlying approximations 
as compared to main stream computational schemes based on density functional theory.
%
The failures of the latter in hexaborides appear to be related to the correlated nature of the B$_6^{2-}$ electronic wavefunctions.

\section*{Methods}

As material model, a \ce{B6^2-} cluster was embedded in a field of $M$ (with $M=\text{Ca, Yb}$) 
and B point charges (see Fig.\:\ref{fig:crystal} and Supplementary Table 2 for the cluster coordinates).
The point charge field was created using the \textsc{Ewald} program
\cite{Derenzo2000,Klintenberg2000}.
%
For the lattice parameters, experimental data were used: 
lattice vectors $a\!=\!4.1537$~{\AA} and an $x$-axis B atom position $x_B\!=\!0.201$ 
for CaB$_6$ \cite{Schmitt2001,Lee2005,Schmidt2015} and parameters of $a$\:=\:$4.1792$\:{\AA} 
and $x_B$\:=\:$0.202$ for YbB$_6$ \cite{Jun2007,YbB6_min_2016}.
To account for covalency effects, an initial charge value of $+1.8$~e was chosen for Ca ions 
in the embedding, which corresponds to an ionic charge of $-0.3$~e for B.
%
In contrast, in \ce{YbB6} initial effective charges of 
$+2.0$\:e for Yb \cite{YbB6_nanba_1993,Gavilano2003,Zhou2015,YbB6_min_2016} and $-0.33$\:e for B
were assumed.
For both embeddings, these charges were slightly adjusted by \textsc{Ewald} 
to ensure convergence of the Madelung sum. 
The final obtained point charge values do not deviate more than $0.01$\:e 
from the initial values.
Charge neutrality was ensured in the overall system by slightly adjusting
a set of about $150$ B point charges at the boundary of the whole array 
of approximately $40 000$ point charges.
To eight $M$ nearest neighbors and to six adjacent \ce{B6} octahedra 
around the reference \ce{B6^2-} quantum cluster (see Fig.\;\ref{fig:crystal}) 
we assigned capped effective core potentials (cECPs);
the pseudopotentials of Kaupp \textit{et al.}~\cite{Kaupp1991}, 
Dolg \textit{et al.} \cite{Dolg1989} and Bergner \textit{et al.}~\cite{Bergner1993} 
were employed for Ca, Yb and B, respectively.
Their main purpose is to prevent electron density from the quantum cluster being 
artificially polarized by nearby point charges.
Consequently, for the quantum cluster, an all-electron correlation-consistent polarized 
valence quadruple zeta (cc-pVQZ) basis set \cite{Dunning1989} was used in conjunction 
with the CASSCF method described above.

\vspace{2em}
\section*{Data Availability Statement}
\noindent
Supplementary data (excitation energies of CaB$_6$ using further methods, cluster and 
point charge geometries) is attached and Orca calculation outputs are deposited at the 
ioChem-BD repository under the 
DOI \href{https://doi.org/10.19061/iochem-bd-6-146}{10.19061/iochem-bd-6-146}.
The datasets generated during and/or analysed during the current study are available 
from the corresponding author on reasonable request.

\section*{Acknowledgments}
\noindent
We thank 
J.~Geck, 
L.~Craco, 
V.~Bisogni  
and P.~Fulde 
for discussions, 
U.~Nitzsche for technical assistance, 
and the German Research Foundation for financial support.

\section*{Contributions}
\noindent
T.P. and L.H. carried out the quantum chemical calculations.
U.K.R. and L.H. planned the project.

\section*{Ethics declarations}
\noindent
The authors declare that there are no competing interests.

\bibliography{refs_may29}

\begin{thebibliography}{52}%
\makeatletter
\providecommand \@ifxundefined [1]{%
 \@ifx{#1\undefined}
}%
\providecommand \@ifnum [1]{%
 \ifnum #1\expandafter \@firstoftwo
 \else \expandafter \@secondoftwo
 \fi
}%
\providecommand \@ifx [1]{%
 \ifx #1\expandafter \@firstoftwo
 \else \expandafter \@secondoftwo
 \fi
}%
\providecommand \natexlab [1]{#1}%
\providecommand \enquote  [1]{``#1''}%
\providecommand \bibnamefont  [1]{#1}%
\providecommand \bibfnamefont [1]{#1}%
\providecommand \citenamefont [1]{#1}%
\providecommand \href@noop [0]{\@secondoftwo}%
\providecommand \href [0]{\begingroup \@sanitize@url \@href}%
\providecommand \@href[1]{\@@startlink{#1}\@@href}%
\providecommand \@@href[1]{\endgroup#1\@@endlink}%
\providecommand \@sanitize@url [0]{\catcode `\\12\catcode `\$12\catcode
  `\&12\catcode `\#12\catcode `\^12\catcode `\_12\catcode `\%12\relax}%
\providecommand \@@startlink[1]{}%
\providecommand \@@endlink[0]{}%
\providecommand \url  [0]{\begingroup\@sanitize@url \@url }%
\providecommand \@url [1]{\endgroup\@href {#1}{\urlprefix }}%
\providecommand \urlprefix  [0]{URL }%
\providecommand \Eprint [0]{\href }%
\providecommand \doibase [0]{https://doi.org/}%
\providecommand \selectlanguage [0]{\@gobble}%
\providecommand \bibinfo  [0]{\@secondoftwo}%
\providecommand \bibfield  [0]{\@secondoftwo}%
\providecommand \translation [1]{[#1]}%
\providecommand \BibitemOpen [0]{}%
\providecommand \bibitemStop [0]{}%
\providecommand \bibitemNoStop [0]{.\EOS\space}%
\providecommand \EOS [0]{\spacefactor3000\relax}%
\providecommand \BibitemShut  [1]{\csname bibitem#1\endcsname}%
\let\auto@bib@innerbib\@empty
\bibitem [{\citenamefont {Inosov}(2021)}]{inosov2021rare}%
  \BibitemOpen
  \bibfield  {author} {\bibinfo {author} {\bibfnamefont {D.}~\bibnamefont
  {Inosov}},\ }\href {https://doi.org/10.1201/9781003146483} {\emph {\bibinfo
  {title} {Rare-Earth Borides}}}\ (\bibinfo  {publisher} {Jenny Stanford
  Publishing},\ \bibinfo {year} {2021})\BibitemShut {NoStop}%
\bibitem [{\citenamefont {Cahill}\ and\ \citenamefont
  {Graeve}(2019)}]{Cahill2019}%
  \BibitemOpen
  \bibfield  {author} {\bibinfo {author} {\bibfnamefont {J.~T.}\ \bibnamefont
  {Cahill}}\ and\ \bibinfo {author} {\bibfnamefont {O.~A.}\ \bibnamefont
  {Graeve}},\ }\bibfield  {title} {\bibinfo {title} {Hexaborides: a review of
  structure, synthesis and processing},\ }\href
  {https://doi.org/10.1016/j.jmrt.2019.09.041} {\bibfield  {journal} {\bibinfo
  {journal} {J. Mater. Res. Technol.}\ }\textbf {\bibinfo {volume} {8}},\
  \bibinfo {pages} {6321} (\bibinfo {year} {2019})}\BibitemShut {NoStop}%
\bibitem [{\citenamefont {Denlinger}\ \emph
  {et~al.}(2002{\natexlab{a}})\citenamefont {Denlinger}, \citenamefont {Clack},
  \citenamefont {Allen}, \citenamefont {Gweon}, \citenamefont {Poirier},
  \citenamefont {Olson}, \citenamefont {Sarrao}, \citenamefont {Bianchi},\ and\
  \citenamefont {Fisk}}]{CaB6_denlinger_2002}%
  \BibitemOpen
  \bibfield  {author} {\bibinfo {author} {\bibfnamefont {J.~D.}\ \bibnamefont
  {Denlinger}}, \bibinfo {author} {\bibfnamefont {J.~A.}\ \bibnamefont
  {Clack}}, \bibinfo {author} {\bibfnamefont {J.~W.}\ \bibnamefont {Allen}},
  \bibinfo {author} {\bibfnamefont {G.-H.}\ \bibnamefont {Gweon}}, \bibinfo
  {author} {\bibfnamefont {D.~M.}\ \bibnamefont {Poirier}}, \bibinfo {author}
  {\bibfnamefont {C.~G.}\ \bibnamefont {Olson}}, \bibinfo {author}
  {\bibfnamefont {J.~L.}\ \bibnamefont {Sarrao}}, \bibinfo {author}
  {\bibfnamefont {A.~D.}\ \bibnamefont {Bianchi}},\ and\ \bibinfo {author}
  {\bibfnamefont {Z.}~\bibnamefont {Fisk}},\ }\bibfield  {title} {\bibinfo
  {title} {Bulk band gaps in divalent hexaborides},\ }\href
  {https://doi.org/10.1103/PhysRevLett.89.157601} {\bibfield  {journal}
  {\bibinfo  {journal} {Phys. Rev. Lett.}\ }\textbf {\bibinfo {volume} {89}},\
  \bibinfo {pages} {157601} (\bibinfo {year} {2002}{\natexlab{a}})}\BibitemShut
  {NoStop}%
\bibitem [{\citenamefont {Lortz}\ \emph {et~al.}(2006)\citenamefont {Lortz},
  \citenamefont {Wang}, \citenamefont {Tutsch}, \citenamefont {Abe},
  \citenamefont {Meingast}, \citenamefont {Popovich}, \citenamefont {Knafo},
  \citenamefont {Shitsevalova}, \citenamefont {Paderno},\ and\ \citenamefont
  {Junod}}]{YB6_junod_2006}%
  \BibitemOpen
  \bibfield  {author} {\bibinfo {author} {\bibfnamefont {R.}~\bibnamefont
  {Lortz}}, \bibinfo {author} {\bibfnamefont {Y.}~\bibnamefont {Wang}},
  \bibinfo {author} {\bibfnamefont {U.}~\bibnamefont {Tutsch}}, \bibinfo
  {author} {\bibfnamefont {S.}~\bibnamefont {Abe}}, \bibinfo {author}
  {\bibfnamefont {C.}~\bibnamefont {Meingast}}, \bibinfo {author}
  {\bibfnamefont {P.}~\bibnamefont {Popovich}}, \bibinfo {author}
  {\bibfnamefont {W.}~\bibnamefont {Knafo}}, \bibinfo {author} {\bibfnamefont
  {N.}~\bibnamefont {Shitsevalova}}, \bibinfo {author} {\bibfnamefont {Y.~B.}\
  \bibnamefont {Paderno}},\ and\ \bibinfo {author} {\bibfnamefont
  {A.}~\bibnamefont {Junod}},\ }\bibfield  {title} {\bibinfo {title}
  {Superconductivity mediated by a soft phonon mode: Specific heat,
  resistivity, thermal expansion, and magnetization of {YB}$_6$},\ }\href
  {https://doi.org/10.1103/PhysRevB.73.024512} {\bibfield  {journal} {\bibinfo
  {journal} {Phys. Rev. B}\ }\textbf {\bibinfo {volume} {73}},\ \bibinfo
  {pages} {024512} (\bibinfo {year} {2006})}\BibitemShut {NoStop}%
\bibitem [{\citenamefont {Swanson}\ \emph {et~al.}(1981)\citenamefont
  {Swanson}, \citenamefont {Gesley},\ and\ \citenamefont
  {Davis}}]{LaB6_swanson_1981}%
  \BibitemOpen
  \bibfield  {author} {\bibinfo {author} {\bibfnamefont {L.~W.}\ \bibnamefont
  {Swanson}}, \bibinfo {author} {\bibfnamefont {M.~A.}\ \bibnamefont
  {Gesley}},\ and\ \bibinfo {author} {\bibfnamefont {P.~R.}\ \bibnamefont
  {Davis}},\ }\bibfield  {title} {\bibinfo {title} {Crystallographic dependence
  of the work function and volatility of {LaB}$_6$},\ }\href
  {https://doi.org/10.1016/0167-2584(81)90558-2} {\bibfield  {journal}
  {\bibinfo  {journal} {Surf. Sci.}\ }\textbf {\bibinfo {volume} {107}},\
  \bibinfo {pages} {263} (\bibinfo {year} {1981})}\BibitemShut {NoStop}%
\bibitem [{\citenamefont {Gavilano}\ \emph {et~al.}(1998)\citenamefont
  {Gavilano}, \citenamefont {Ambrosini}, \citenamefont {Vonlanthen},
  \citenamefont {Ott}, \citenamefont {Young},\ and\ \citenamefont
  {Fisk}}]{EuB6_fisk_1998}%
  \BibitemOpen
  \bibfield  {author} {\bibinfo {author} {\bibfnamefont {J.~L.}\ \bibnamefont
  {Gavilano}}, \bibinfo {author} {\bibfnamefont {B.}~\bibnamefont {Ambrosini}},
  \bibinfo {author} {\bibfnamefont {P.}~\bibnamefont {Vonlanthen}}, \bibinfo
  {author} {\bibfnamefont {H.~R.}\ \bibnamefont {Ott}}, \bibinfo {author}
  {\bibfnamefont {D.~P.}\ \bibnamefont {Young}},\ and\ \bibinfo {author}
  {\bibfnamefont {Z.}~\bibnamefont {Fisk}},\ }\bibfield  {title} {\bibinfo
  {title} {Low temperature nuclear magnetic resonance studies of {EuB}$_6$},\
  }\href {https://doi.org/10.1103/PhysRevLett.81.5648} {\bibfield  {journal}
  {\bibinfo  {journal} {Phys. Rev. Lett.}\ }\textbf {\bibinfo {volume} {81}},\
  \bibinfo {pages} {25} (\bibinfo {year} {1998})}\BibitemShut {NoStop}%
\bibitem [{\citenamefont {Cameron}\ \emph {et~al.}(2016)\citenamefont
  {Cameron}, \citenamefont {Friemel},\ and\ \citenamefont
  {Inosov}}]{CeB6_inosov_2016}%
  \BibitemOpen
  \bibfield  {author} {\bibinfo {author} {\bibfnamefont {A.~S.}\ \bibnamefont
  {Cameron}}, \bibinfo {author} {\bibfnamefont {G.}~\bibnamefont {Friemel}},\
  and\ \bibinfo {author} {\bibfnamefont {D.~S.}\ \bibnamefont {Inosov}},\
  }\bibfield  {title} {\bibinfo {title} {Multipolar phases and magnetically
  hidden order: review of the heavy-fermion compound
  {Ce}$_{1-x}${La}$_x${B}$_6$},\ }\href
  {https://doi.org/10.1088/0034-4885/79/6/066502} {\bibfield  {journal}
  {\bibinfo  {journal} {Rep. Prog. Phys.}\ }\textbf {\bibinfo {volume} {79}},\
  \bibinfo {pages} {066502} (\bibinfo {year} {2016})}\BibitemShut {NoStop}%
\bibitem [{\citenamefont {Thalmeier}\ \emph {et~al.}(2019)\citenamefont
  {Thalmeier}, \citenamefont {Akbari},\ and\ \citenamefont
  {Shiina}}]{CeB6_thalmeier_2019}%
  \BibitemOpen
  \bibfield  {author} {\bibinfo {author} {\bibfnamefont {P.}~\bibnamefont
  {Thalmeier}}, \bibinfo {author} {\bibfnamefont {A.}~\bibnamefont {Akbari}},\
  and\ \bibinfo {author} {\bibfnamefont {R.}~\bibnamefont {Shiina}},\
  }\bibfield  {title} {\bibinfo {title} {Multipolar order and excitations in
  rare-earth boride {Kondo} systems},\ }in\ \href
  {https://arxiv.org/abs/1907.10967} {\emph {\bibinfo {booktitle} {Rare-Earth
  Borides}}},\ \bibinfo {editor} {edited by\ \bibinfo {editor} {\bibfnamefont
  {D.}~\bibnamefont {Inosov}}}\ (\bibinfo  {publisher} {Jenny Stanford
  Publishing},\ \bibinfo {year} {2019})\ Chap.~\bibinfo {chapter}
  {8}\BibitemShut {NoStop}%
\bibitem [{\citenamefont {Li}\ \emph {et~al.}(2020)\citenamefont {Li},
  \citenamefont {Sun}, \citenamefont {Kurdak},\ and\ \citenamefont
  {Allen}}]{SmB6_allen_2020}%
  \BibitemOpen
  \bibfield  {author} {\bibinfo {author} {\bibfnamefont {L.}~\bibnamefont
  {Li}}, \bibinfo {author} {\bibfnamefont {K.}~\bibnamefont {Sun}}, \bibinfo
  {author} {\bibfnamefont {C.}~\bibnamefont {Kurdak}},\ and\ \bibinfo {author}
  {\bibfnamefont {J.~W.}\ \bibnamefont {Allen}},\ }\bibfield  {title} {\bibinfo
  {title} {Emergent mystery in the {K}ondo insulator samarium hexaboride},\
  }\href {https://doi.org/10.1038/s42254-020-0210-8} {\bibfield  {journal}
  {\bibinfo  {journal} {Nat. Rev. Phys.}\ }\textbf {\bibinfo {volume} {2}},\
  \bibinfo {pages} {463} (\bibinfo {year} {2020})}\BibitemShut {NoStop}%
\bibitem [{\citenamefont {Hasegawa}\ and\ \citenamefont
  {Yanase}(1979)}]{CaB6_hasegawa_1979}%
  \BibitemOpen
  \bibfield  {author} {\bibinfo {author} {\bibfnamefont {A.}~\bibnamefont
  {Hasegawa}}\ and\ \bibinfo {author} {\bibfnamefont {A.}~\bibnamefont
  {Yanase}},\ }\bibfield  {title} {\bibinfo {title} {Electronic structure of
  {CaB}$_6$},\ }\href {https://doi.org/10.1088/0022-3719/12/24/014} {\bibfield
  {journal} {\bibinfo  {journal} {J. Phys. C: Solid State Phys.}\ }\textbf
  {\bibinfo {volume} {12}},\ \bibinfo {pages} {5431} (\bibinfo {year}
  {1979})}\BibitemShut {NoStop}%
\bibitem [{\citenamefont {Massidda}\ \emph {et~al.}(1997)\citenamefont
  {Massidda}, \citenamefont {Continenza}, \citenamefont {Pascale},\ and\
  \citenamefont {Monnier}}]{CaB6_massidda_1997}%
  \BibitemOpen
  \bibfield  {author} {\bibinfo {author} {\bibfnamefont {S.}~\bibnamefont
  {Massidda}}, \bibinfo {author} {\bibfnamefont {A.}~\bibnamefont
  {Continenza}}, \bibinfo {author} {\bibfnamefont {T.~M.~D.}\ \bibnamefont
  {Pascale}},\ and\ \bibinfo {author} {\bibfnamefont {R.}~\bibnamefont
  {Monnier}},\ }\bibfield  {title} {\bibinfo {title} {Electronic structure of
  divalent hexaborides},\ }\href {https://doi.org/10.1007/S002570050267}
  {\bibfield  {journal} {\bibinfo  {journal} {Z. Phys. B}\ }\textbf {\bibinfo
  {volume} {102}},\ \bibinfo {pages} {83} (\bibinfo {year} {1997})}\BibitemShut
  {NoStop}%
\bibitem [{\citenamefont {Rodriguez}\ \emph {et~al.}(2000)\citenamefont
  {Rodriguez}, \citenamefont {Weht},\ and\ \citenamefont
  {Pickett}}]{CaB6_pickett_2000}%
  \BibitemOpen
  \bibfield  {author} {\bibinfo {author} {\bibfnamefont {C.~O.}\ \bibnamefont
  {Rodriguez}}, \bibinfo {author} {\bibfnamefont {R.}~\bibnamefont {Weht}},\
  and\ \bibinfo {author} {\bibfnamefont {W.~E.}\ \bibnamefont {Pickett}},\
  }\bibfield  {title} {\bibinfo {title} {Electronic fine structure in the
  electron-hole plasma in {SrB}$_6$},\ }\href
  {https://doi.org/10.1103/PhysRevLett.84.3903} {\bibfield  {journal} {\bibinfo
   {journal} {Phys. Rev. Lett.}\ }\textbf {\bibinfo {volume} {84}},\ \bibinfo
  {pages} {3903} (\bibinfo {year} {2000})}\BibitemShut {NoStop}%
\bibitem [{\citenamefont {Kino}\ \emph
  {et~al.}(2002{\natexlab{a}})\citenamefont {Kino}, \citenamefont
  {Aryasetiawan}, \citenamefont {Terakura},\ and\ \citenamefont
  {Miyake}}]{CaB6_aryasetiawan_2002}%
  \BibitemOpen
  \bibfield  {author} {\bibinfo {author} {\bibfnamefont {H.}~\bibnamefont
  {Kino}}, \bibinfo {author} {\bibfnamefont {F.}~\bibnamefont {Aryasetiawan}},
  \bibinfo {author} {\bibfnamefont {K.}~\bibnamefont {Terakura}},\ and\
  \bibinfo {author} {\bibfnamefont {T.}~\bibnamefont {Miyake}},\ }\bibfield
  {title} {\bibinfo {title} {Abnormal quasiparticle shifts in {CaB}$_6$},\
  }\href {https://doi.org/10.1103/PhysRevB.66.121103} {\bibfield  {journal}
  {\bibinfo  {journal} {Phys. Rev. B}\ }\textbf {\bibinfo {volume} {66}},\
  \bibinfo {pages} {121103(R)} (\bibinfo {year}
  {2002}{\natexlab{a}})}\BibitemShut {NoStop}%
\bibitem [{\citenamefont {Kino}\ \emph
  {et~al.}(2002{\natexlab{b}})\citenamefont {Kino}, \citenamefont
  {Aryasetiawan}, \citenamefont {{van Schilfgaarde}}, \citenamefont {Kotani},
  \citenamefont {Miyake},\ and\ \citenamefont {Terakura}}]{CaB6_kino_2002}%
  \BibitemOpen
  \bibfield  {author} {\bibinfo {author} {\bibfnamefont {H.}~\bibnamefont
  {Kino}}, \bibinfo {author} {\bibfnamefont {F.}~\bibnamefont {Aryasetiawan}},
  \bibinfo {author} {\bibfnamefont {M.}~\bibnamefont {{van Schilfgaarde}}},
  \bibinfo {author} {\bibfnamefont {T.}~\bibnamefont {Kotani}}, \bibinfo
  {author} {\bibfnamefont {T.}~\bibnamefont {Miyake}},\ and\ \bibinfo {author}
  {\bibfnamefont {K.}~\bibnamefont {Terakura}},\ }\bibfield  {title} {\bibinfo
  {title} {{GW} quasiparticle band structure of {CaB}$_6$},\ }\href
  {https://doi.org/https://doi.org/10.1016/S0022-3697(02)00118-X} {\bibfield
  {journal} {\bibinfo  {journal} {J. Phys. Chem. Solids}\ }\textbf {\bibinfo
  {volume} {63}},\ \bibinfo {pages} {1595} (\bibinfo {year}
  {2002}{\natexlab{b}})}\BibitemShut {NoStop}%
\bibitem [{\citenamefont {Neupane}\ \emph {et~al.}(2015)\citenamefont
  {Neupane}, \citenamefont {Xu}, \citenamefont {Alidoust}, \citenamefont
  {Bian}, \citenamefont {Kim}, \citenamefont {Liu}, \citenamefont {Belopolski},
  \citenamefont {Chang}, \citenamefont {Jeng}, \citenamefont {Durakiewicz},
  \citenamefont {Lin}, \citenamefont {Bansil}, \citenamefont {Fisk},\ and\
  \citenamefont {Hasan}}]{YbB6_hasan_2015}%
  \BibitemOpen
  \bibfield  {author} {\bibinfo {author} {\bibfnamefont {M.}~\bibnamefont
  {Neupane}}, \bibinfo {author} {\bibfnamefont {S.-Y.}\ \bibnamefont {Xu}},
  \bibinfo {author} {\bibfnamefont {N.}~\bibnamefont {Alidoust}}, \bibinfo
  {author} {\bibfnamefont {G.}~\bibnamefont {Bian}}, \bibinfo {author}
  {\bibfnamefont {D.~J.}\ \bibnamefont {Kim}}, \bibinfo {author} {\bibfnamefont
  {C.}~\bibnamefont {Liu}}, \bibinfo {author} {\bibfnamefont {I.}~\bibnamefont
  {Belopolski}}, \bibinfo {author} {\bibfnamefont {T.-R.}\ \bibnamefont
  {Chang}}, \bibinfo {author} {\bibfnamefont {H.-T.}\ \bibnamefont {Jeng}},
  \bibinfo {author} {\bibfnamefont {T.}~\bibnamefont {Durakiewicz}}, \bibinfo
  {author} {\bibfnamefont {H.}~\bibnamefont {Lin}}, \bibinfo {author}
  {\bibfnamefont {A.}~\bibnamefont {Bansil}}, \bibinfo {author} {\bibfnamefont
  {Z.}~\bibnamefont {Fisk}},\ and\ \bibinfo {author} {\bibfnamefont {M.~Z.}\
  \bibnamefont {Hasan}},\ }\bibfield  {title} {\bibinfo {title}
  {Non-{Kondo}-like electronic structure in the correlated rare-earth
  hexaboride {YbB}$_6$},\ }\href
  {https://doi.org/10.1103/PhysRevLett.114.016403} {\bibfield  {journal}
  {\bibinfo  {journal} {Phys. Rev. Lett.}\ }\textbf {\bibinfo {volume} {114}},\
  \bibinfo {pages} {016403} (\bibinfo {year} {2015})}\BibitemShut {NoStop}%
\bibitem [{\citenamefont {Zhang}\ \emph {et~al.}(2020)\citenamefont {Zhang},
  \citenamefont {Li}, \citenamefont {Sun}, \citenamefont {Qin}, \citenamefont
  {Kang}, \citenamefont {Yao}, \citenamefont {Weng}, \citenamefont {Mo},
  \citenamefont {Li}, \citenamefont {Liu}, \citenamefont {Yang},\ and\
  \citenamefont {Chen}}]{YbB6_chen_2020}%
  \BibitemOpen
  \bibfield  {author} {\bibinfo {author} {\bibfnamefont {T.}~\bibnamefont
  {Zhang}}, \bibinfo {author} {\bibfnamefont {G.}~\bibnamefont {Li}}, \bibinfo
  {author} {\bibfnamefont {S.~C.}\ \bibnamefont {Sun}}, \bibinfo {author}
  {\bibfnamefont {N.}~\bibnamefont {Qin}}, \bibinfo {author} {\bibfnamefont
  {L.}~\bibnamefont {Kang}}, \bibinfo {author} {\bibfnamefont {S.~H.}\
  \bibnamefont {Yao}}, \bibinfo {author} {\bibfnamefont {H.~M.}\ \bibnamefont
  {Weng}}, \bibinfo {author} {\bibfnamefont {S.~K.}\ \bibnamefont {Mo}},
  \bibinfo {author} {\bibfnamefont {L.}~\bibnamefont {Li}}, \bibinfo {author}
  {\bibfnamefont {Z.~K.}\ \bibnamefont {Liu}}, \bibinfo {author} {\bibfnamefont
  {L.~X.}\ \bibnamefont {Yang}},\ and\ \bibinfo {author} {\bibfnamefont
  {Y.~L.}\ \bibnamefont {Chen}},\ }\bibfield  {title} {\bibinfo {title}
  {Electronic structure of correlated topological insulator candidate {YbB}$_6$
  studied by photoemission and quantum oscillation},\ }\href
  {https://doi.org/10.1088/1674-1056/ab6206} {\bibfield  {journal} {\bibinfo
  {journal} {Chin. Phys. B}\ }\textbf {\bibinfo {volume} {29}},\ \bibinfo
  {pages} {017304} (\bibinfo {year} {2020})}\BibitemShut {NoStop}%
\bibitem [{\citenamefont {Kang}\ \emph {et~al.}(2016)\citenamefont {Kang},
  \citenamefont {Denlinger}, \citenamefont {Allen}, \citenamefont {Min},
  \citenamefont {Reinert}, \citenamefont {Kang}, \citenamefont {Cho},
  \citenamefont {Kang}, \citenamefont {Shim},\ and\ \citenamefont
  {Min}}]{YbB6_min_2016}%
  \BibitemOpen
  \bibfield  {author} {\bibinfo {author} {\bibfnamefont {C.-J.}\ \bibnamefont
  {Kang}}, \bibinfo {author} {\bibfnamefont {J.~D.}\ \bibnamefont {Denlinger}},
  \bibinfo {author} {\bibfnamefont {J.~W.}\ \bibnamefont {Allen}}, \bibinfo
  {author} {\bibfnamefont {C.-H.}\ \bibnamefont {Min}}, \bibinfo {author}
  {\bibfnamefont {F.}~\bibnamefont {Reinert}}, \bibinfo {author} {\bibfnamefont
  {B.~Y.}\ \bibnamefont {Kang}}, \bibinfo {author} {\bibfnamefont {B.~K.}\
  \bibnamefont {Cho}}, \bibinfo {author} {\bibfnamefont {J.-S.}\ \bibnamefont
  {Kang}}, \bibinfo {author} {\bibfnamefont {J.~H.}\ \bibnamefont {Shim}},\
  and\ \bibinfo {author} {\bibfnamefont {B.~I.}\ \bibnamefont {Min}},\
  }\bibfield  {title} {\bibinfo {title} {Electronic structure of {YbB}$_6$: Is
  it a topological insulator or not?},\ }\href
  {https://doi.org/10.1103/PhysRevLett.116.116401} {\bibfield  {journal}
  {\bibinfo  {journal} {Phys. Rev. Lett.}\ }\textbf {\bibinfo {volume} {116}},\
  \bibinfo {pages} {116401} (\bibinfo {year} {2016})}\BibitemShut {NoStop}%
\bibitem [{\citenamefont {Sundermann}\ \emph {et~al.}(2018)\citenamefont
  {Sundermann}, \citenamefont {Yava}, \citenamefont {Chen}, \citenamefont
  {Kim}, \citenamefont {Fisk}, \citenamefont {Kasinathan}, \citenamefont
  {Haverkort}, \citenamefont {Thalmeier}, \citenamefont {Severing},\ and\
  \citenamefont {Tjeng}}]{SmB6_tjeng_2018}%
  \BibitemOpen
  \bibfield  {author} {\bibinfo {author} {\bibfnamefont {M.}~\bibnamefont
  {Sundermann}}, \bibinfo {author} {\bibfnamefont {H.}~\bibnamefont {Yava}},
  \bibinfo {author} {\bibfnamefont {K.}~\bibnamefont {Chen}}, \bibinfo {author}
  {\bibfnamefont {D.~J.}\ \bibnamefont {Kim}}, \bibinfo {author} {\bibfnamefont
  {Z.}~\bibnamefont {Fisk}}, \bibinfo {author} {\bibfnamefont {D.}~\bibnamefont
  {Kasinathan}}, \bibinfo {author} {\bibfnamefont {M.~W.}\ \bibnamefont
  {Haverkort}}, \bibinfo {author} {\bibfnamefont {P.}~\bibnamefont
  {Thalmeier}}, \bibinfo {author} {\bibfnamefont {A.}~\bibnamefont
  {Severing}},\ and\ \bibinfo {author} {\bibfnamefont {L.~H.}\ \bibnamefont
  {Tjeng}},\ }\bibfield  {title} {\bibinfo {title} {4$f$ crystal field ground
  state of the strongly correlated topological insulator {SmB}$_6$},\ }\href
  {https://doi.org/10.1103/PhysRevLett.120.016402} {\bibfield  {journal}
  {\bibinfo  {journal} {Phys. Rev. Lett.}\ }\textbf {\bibinfo {volume} {120}},\
  \bibinfo {pages} {016402} (\bibinfo {year} {2018})}\BibitemShut {NoStop}%
\bibitem [{\citenamefont {Amorese}\ \emph {et~al.}(2019)\citenamefont
  {Amorese}, \citenamefont {Stockert}, \citenamefont {Kummer}, \citenamefont
  {Brookes}, \citenamefont {Kim}, \citenamefont {Fisk}, \citenamefont
  {Haverkort}, \citenamefont {Thalmeier}, \citenamefont {Tjeng},\ and\
  \citenamefont {Severing}}]{SmB6_tjeng_2019}%
  \BibitemOpen
  \bibfield  {author} {\bibinfo {author} {\bibfnamefont {A.}~\bibnamefont
  {Amorese}}, \bibinfo {author} {\bibfnamefont {O.}~\bibnamefont {Stockert}},
  \bibinfo {author} {\bibfnamefont {K.}~\bibnamefont {Kummer}}, \bibinfo
  {author} {\bibfnamefont {N.~B.}\ \bibnamefont {Brookes}}, \bibinfo {author}
  {\bibfnamefont {D.~J.}\ \bibnamefont {Kim}}, \bibinfo {author} {\bibfnamefont
  {Z.}~\bibnamefont {Fisk}}, \bibinfo {author} {\bibfnamefont {M.~W.}\
  \bibnamefont {Haverkort}}, \bibinfo {author} {\bibfnamefont {P.}~\bibnamefont
  {Thalmeier}}, \bibinfo {author} {\bibfnamefont {L.~H.}\ \bibnamefont
  {Tjeng}},\ and\ \bibinfo {author} {\bibfnamefont {A.}~\bibnamefont
  {Severing}},\ }\bibfield  {title} {\bibinfo {title} {Resonant inelastic x-ray
  scattering investigation of the crystal-field splitting of {Sm}$^{3+}$ in
  {SmB}$_6$},\ }\href {https://doi.org/10.1103/PhysRevB.100.241107} {\bibfield
  {journal} {\bibinfo  {journal} {Phys. Rev. B}\ }\textbf {\bibinfo {volume}
  {100}},\ \bibinfo {pages} {241107(R)} (\bibinfo {year} {2019})}\BibitemShut
  {NoStop}%
\bibitem [{\citenamefont {Denlinger}\ \emph
  {et~al.}(2002{\natexlab{b}})\citenamefont {Denlinger}, \citenamefont {Gweon},
  \citenamefont {Allen}, \citenamefont {Bianchi},\ and\ \citenamefont
  {Fisk}}]{YbB6_denlinger_2002}%
  \BibitemOpen
  \bibfield  {author} {\bibinfo {author} {\bibfnamefont {J.~D.}\ \bibnamefont
  {Denlinger}}, \bibinfo {author} {\bibfnamefont {G.-H.}\ \bibnamefont
  {Gweon}}, \bibinfo {author} {\bibfnamefont {J.~W.}\ \bibnamefont {Allen}},
  \bibinfo {author} {\bibfnamefont {A.~D.}\ \bibnamefont {Bianchi}},\ and\
  \bibinfo {author} {\bibfnamefont {Z.}~\bibnamefont {Fisk}},\ }\bibfield
  {title} {\bibinfo {title} {Bulk band gaps in divalent hexaborides: A soft
  x-ray emission study},\ }\href {https://doi.org/10.1142/S0218625X0200372X}
  {\bibfield  {journal} {\bibinfo  {journal} {Surf. Rev. Lett.}\ }\textbf
  {\bibinfo {volume} {09}},\ \bibinfo {pages} {1309} (\bibinfo {year}
  {2002}{\natexlab{b}})}\BibitemShut {NoStop}%
\bibitem [{\citenamefont {Hozoi}\ \emph {et~al.}(2008)\citenamefont {Hozoi},
  \citenamefont {Laad},\ and\ \citenamefont {Fulde}}]{CuO_hozoi_08}%
  \BibitemOpen
  \bibfield  {author} {\bibinfo {author} {\bibfnamefont {L.}~\bibnamefont
  {Hozoi}}, \bibinfo {author} {\bibfnamefont {M.~S.}\ \bibnamefont {Laad}},\
  and\ \bibinfo {author} {\bibfnamefont {P.}~\bibnamefont {Fulde}},\ }\bibfield
   {title} {\bibinfo {title} {Fermiology of cuprates from first principles:
  From small pockets to the {Luttinger} {Fermi} surface},\ }\href
  {https://doi.org/10.1103/PhysRevB.78.165107} {\bibfield  {journal} {\bibinfo
  {journal} {Phys. Rev. B}\ }\textbf {\bibinfo {volume} {78}},\ \bibinfo
  {pages} {165107} (\bibinfo {year} {2008})}\BibitemShut {NoStop}%
\bibitem [{\citenamefont {Hozoi}\ \emph {et~al.}(2007)\citenamefont {Hozoi},
  \citenamefont {Birkenheuer}, \citenamefont {Fulde}, \citenamefont
  {Mitrushchenkov},\ and\ \citenamefont {Stoll}}]{MgO_hozoi_07}%
  \BibitemOpen
  \bibfield  {author} {\bibinfo {author} {\bibfnamefont {L.}~\bibnamefont
  {Hozoi}}, \bibinfo {author} {\bibfnamefont {U.}~\bibnamefont {Birkenheuer}},
  \bibinfo {author} {\bibfnamefont {P.}~\bibnamefont {Fulde}}, \bibinfo
  {author} {\bibfnamefont {A.~O.}\ \bibnamefont {Mitrushchenkov}},\ and\
  \bibinfo {author} {\bibfnamefont {H.}~\bibnamefont {Stoll}},\ }\bibfield
  {title} {\bibinfo {title} {Ab initio wave function-based methods for excited
  states in solids: correlation corrections to the band structure of ionic
  oxides},\ }\href {https://doi.org/10.1103/PhysRevB.76.085109} {\bibfield
  {journal} {\bibinfo  {journal} {Phys. Rev. B}\ }\textbf {\bibinfo {volume}
  {76}},\ \bibinfo {pages} {085109} (\bibinfo {year} {2007})}\BibitemShut
  {NoStop}%
\bibitem [{\citenamefont {Stoyanova}\ \emph {et~al.}(2014)\citenamefont
  {Stoyanova}, \citenamefont {Mitrushchenkov}, \citenamefont {Hozoi},
  \citenamefont {Stoll},\ and\ \citenamefont {Fulde}}]{diamond_alex_14}%
  \BibitemOpen
  \bibfield  {author} {\bibinfo {author} {\bibfnamefont {A.}~\bibnamefont
  {Stoyanova}}, \bibinfo {author} {\bibfnamefont {A.~O.}\ \bibnamefont
  {Mitrushchenkov}}, \bibinfo {author} {\bibfnamefont {L.}~\bibnamefont
  {Hozoi}}, \bibinfo {author} {\bibfnamefont {H.}~\bibnamefont {Stoll}},\ and\
  \bibinfo {author} {\bibfnamefont {P.}~\bibnamefont {Fulde}},\ }\bibfield
  {title} {\bibinfo {title} {Electron correlation effects in diamond: {A}
  wave-function quantum-chemistry study of the quasiparticle band structure},\
  }\href {https://doi.org/10.1103/PhysRevB.89.235121} {\bibfield  {journal}
  {\bibinfo  {journal} {Phys. Rev. B}\ }\textbf {\bibinfo {volume} {89}},\
  \bibinfo {pages} {235121} (\bibinfo {year} {2014})}\BibitemShut {NoStop}%
\bibitem [{\citenamefont {Tromp}\ \emph {et~al.}(2001)\citenamefont {Tromp},
  \citenamefont {van Gelderen}, \citenamefont {Kelly}, \citenamefont {Brocks},\
  and\ \citenamefont {Bobbert}}]{CaB6_bobbert_2001}%
  \BibitemOpen
  \bibfield  {author} {\bibinfo {author} {\bibfnamefont {H.~J.}\ \bibnamefont
  {Tromp}}, \bibinfo {author} {\bibfnamefont {P.}~\bibnamefont {van Gelderen}},
  \bibinfo {author} {\bibfnamefont {P.~J.}\ \bibnamefont {Kelly}}, \bibinfo
  {author} {\bibfnamefont {G.}~\bibnamefont {Brocks}},\ and\ \bibinfo {author}
  {\bibfnamefont {P.~A.}\ \bibnamefont {Bobbert}},\ }\bibfield  {title}
  {\bibinfo {title} {{CaB}$_6$: A new semiconducting material for spin
  electronics},\ }\href {https://doi.org/10.1103/PhysRevLett.87.016401}
  {\bibfield  {journal} {\bibinfo  {journal} {Phys. Rev. Lett.}\ }\textbf
  {\bibinfo {volume} {87}},\ \bibinfo {pages} {016401} (\bibinfo {year}
  {2001})}\BibitemShut {NoStop}%
\bibitem [{\citenamefont {Helgaker}\ \emph {et~al.}(2000)\citenamefont
  {Helgaker}, \citenamefont {J{\o}rgensen},\ and\ \citenamefont
  {Olsen}}]{Helgaker2012}%
  \BibitemOpen
  \bibfield  {author} {\bibinfo {author} {\bibfnamefont {T.}~\bibnamefont
  {Helgaker}}, \bibinfo {author} {\bibfnamefont {P.}~\bibnamefont
  {J{\o}rgensen}},\ and\ \bibinfo {author} {\bibfnamefont {J.}~\bibnamefont
  {Olsen}},\ }\href@noop {} {\emph {\bibinfo {title} {Molecular
  Electronic-Structure Theory}}}\ (\bibinfo  {publisher} {Wiley VCH,
  Chichester},\ \bibinfo {year} {2000})\BibitemShut {NoStop}%
\bibitem [{\citenamefont {Neese}\ \emph {et~al.}(2020)\citenamefont {Neese},
  \citenamefont {Wennmohs}, \citenamefont {Becker},\ and\ \citenamefont
  {Riplinger}}]{Neese2020}%
  \BibitemOpen
  \bibfield  {author} {\bibinfo {author} {\bibfnamefont {F.}~\bibnamefont
  {Neese}}, \bibinfo {author} {\bibfnamefont {F.}~\bibnamefont {Wennmohs}},
  \bibinfo {author} {\bibfnamefont {U.}~\bibnamefont {Becker}},\ and\ \bibinfo
  {author} {\bibfnamefont {C.}~\bibnamefont {Riplinger}},\ }\bibfield  {title}
  {\bibinfo {title} {The {ORCA} quantum chemistry program package},\ }\href
  {https://doi.org/10.1063/5.0004608} {\bibfield  {journal} {\bibinfo
  {journal} {J. Chem. Phys.}\ }\textbf {\bibinfo {volume} {152}},\ \bibinfo
  {pages} {224108} (\bibinfo {year} {2020})}\BibitemShut {NoStop}%
\bibitem [{\citenamefont {Janssen}\ and\ \citenamefont
  {Nieuwpoort}(1988)}]{NiO_wim_88}%
  \BibitemOpen
  \bibfield  {author} {\bibinfo {author} {\bibfnamefont {G.~J.~M.}\
  \bibnamefont {Janssen}}\ and\ \bibinfo {author} {\bibfnamefont {W.~C.}\
  \bibnamefont {Nieuwpoort}},\ }\bibfield  {title} {\bibinfo {title} {Band gap
  in {NiO}: A cluster study},\ }\href
  {https://doi.org/10.1103/PhysRevB.38.3449} {\bibfield  {journal} {\bibinfo
  {journal} {Phys. Rev. B}\ }\textbf {\bibinfo {volume} {38}},\ \bibinfo
  {pages} {3449} (\bibinfo {year} {1988})}\BibitemShut {NoStop}%
\bibitem [{\citenamefont {Martin}(1993)}]{CuO_martin_93}%
  \BibitemOpen
  \bibfield  {author} {\bibinfo {author} {\bibfnamefont {R.~L.}\ \bibnamefont
  {Martin}},\ }\bibfield  {title} {\bibinfo {title} {Cluster studies of
  {La}$_2${CuO}$_4$: A mapping onto the {Pariser-Parr-Pople} ({PPP}) model},\
  }\href {https://doi.org/10.1063/1.464476} {\bibfield  {journal} {\bibinfo
  {journal} {J. Chem. Phys.}\ }\textbf {\bibinfo {volume} {98}},\ \bibinfo
  {pages} {8691} (\bibinfo {year} {1993})}\BibitemShut {NoStop}%
\bibitem [{\citenamefont {Bogdanov}\ \emph {et~al.}(2018)\citenamefont
  {Bogdanov}, \citenamefont {Manni}, \citenamefont {Sharma}, \citenamefont
  {Gunnarsson},\ and\ \citenamefont {Alavi}}]{CuO_niko_18}%
  \BibitemOpen
  \bibfield  {author} {\bibinfo {author} {\bibfnamefont {N.~A.}\ \bibnamefont
  {Bogdanov}}, \bibinfo {author} {\bibfnamefont {G.~L.}\ \bibnamefont {Manni}},
  \bibinfo {author} {\bibfnamefont {S.}~\bibnamefont {Sharma}}, \bibinfo
  {author} {\bibfnamefont {O.}~\bibnamefont {Gunnarsson}},\ and\ \bibinfo
  {author} {\bibfnamefont {A.}~\bibnamefont {Alavi}},\ }\bibfield  {title}
  {\bibinfo {title} {New superexchange paths due to breathing-enhanced hopping
  in corner-sharing cuprates},\ }\href {https://arxiv.org/abs/1803.07026}
  {\bibfield  {journal} {\bibinfo  {journal} {arXiv:1803.07026}\ } (\bibinfo
  {year} {2018})}\BibitemShut {NoStop}%
\bibitem [{\citenamefont {Wehling}\ \emph {et~al.}(2011)\citenamefont
  {Wehling}, \citenamefont {\ifmmode \mbox{\c{S}}\else \c{S}\fi{}a\ifmmode
  \mbox{\c{s}}\else \c{s}\fi{}\ifmmode \imath \else \i
  \fi{}o\ifmmode~\breve{g}\else \u{g}\fi{}lu}, \citenamefont {Friedrich},
  \citenamefont {Lichtenstein}, \citenamefont {Katsnelson},\ and\ \citenamefont
  {Bl\"ugel}}]{Wehling2011}%
  \BibitemOpen
  \bibfield  {author} {\bibinfo {author} {\bibfnamefont {T.~O.}\ \bibnamefont
  {Wehling}}, \bibinfo {author} {\bibfnamefont {E.}~\bibnamefont {\ifmmode
  \mbox{\c{S}}\else \c{S}\fi{}a\ifmmode \mbox{\c{s}}\else \c{s}\fi{}\ifmmode
  \imath \else \i \fi{}o\ifmmode~\breve{g}\else \u{g}\fi{}lu}}, \bibinfo
  {author} {\bibfnamefont {C.}~\bibnamefont {Friedrich}}, \bibinfo {author}
  {\bibfnamefont {A.~I.}\ \bibnamefont {Lichtenstein}}, \bibinfo {author}
  {\bibfnamefont {M.~I.}\ \bibnamefont {Katsnelson}},\ and\ \bibinfo {author}
  {\bibfnamefont {S.}~\bibnamefont {Bl\"ugel}},\ }\bibfield  {title} {\bibinfo
  {title} {Strength of effective {Coulomb} interactions in graphene and
  graphite},\ }\href {https://doi.org/10.1103/PhysRevLett.106.236805}
  {\bibfield  {journal} {\bibinfo  {journal} {Phys. Rev. Lett.}\ }\textbf
  {\bibinfo {volume} {106}},\ \bibinfo {pages} {236805} (\bibinfo {year}
  {2011})}\BibitemShut {NoStop}%
\bibitem [{\citenamefont {Craco}\ \emph {et~al.}(2017)\citenamefont {Craco},
  \citenamefont {da~Silva~Pereira},\ and\ \citenamefont
  {Leoni}}]{P_3p_luis_2017}%
  \BibitemOpen
  \bibfield  {author} {\bibinfo {author} {\bibfnamefont {L.}~\bibnamefont
  {Craco}}, \bibinfo {author} {\bibfnamefont {T.~A.}\ \bibnamefont
  {da~Silva~Pereira}},\ and\ \bibinfo {author} {\bibfnamefont {S.}~\bibnamefont
  {Leoni}},\ }\bibfield  {title} {\bibinfo {title} {Electronic structure and
  thermoelectric transport of black phosphorus},\ }\href
  {https://doi.org/https://doi.org/10.1103/PhysRevB.96.075118} {\bibfield
  {journal} {\bibinfo  {journal} {Phys. Rev. B}\ }\textbf {\bibinfo {volume}
  {96}},\ \bibinfo {pages} {075118} (\bibinfo {year} {2017})}\BibitemShut
  {NoStop}%
\bibitem [{\citenamefont {Jensen}(2007)}]{Jensen2007}%
  \BibitemOpen
  \bibfield  {author} {\bibinfo {author} {\bibfnamefont {F.}~\bibnamefont
  {Jensen}},\ }\href@noop {} {\emph {\bibinfo {title} {Introduction to
  Computational Chemistry}}},\ \bibinfo {edition} {2nd}\ ed.\ (\bibinfo
  {publisher} {Wiley \& Sons Ltd.},\ \bibinfo {year} {2007})\BibitemShut
  {NoStop}%
\bibitem [{\citenamefont {Hozoi}\ \emph {et~al.}(2009)\citenamefont {Hozoi},
  \citenamefont {Birkenheuer}, \citenamefont {Stoll},\ and\ \citenamefont
  {Fulde}}]{lacoo3_hozoi_09}%
  \BibitemOpen
  \bibfield  {author} {\bibinfo {author} {\bibfnamefont {L.}~\bibnamefont
  {Hozoi}}, \bibinfo {author} {\bibfnamefont {U.}~\bibnamefont {Birkenheuer}},
  \bibinfo {author} {\bibfnamefont {H.}~\bibnamefont {Stoll}},\ and\ \bibinfo
  {author} {\bibfnamefont {P.}~\bibnamefont {Fulde}},\ }\bibfield  {title}
  {\bibinfo {title} {Spin-state transition and spin-polaron physics in cobalt
  oxide perovskites: ab initio approach based on quantum chemical methods},\
  }\href {https://doi.org/10.1088/1367-2630/11/2/023023} {\bibfield  {journal}
  {\bibinfo  {journal} {New J. Phys.}\ }\textbf {\bibinfo {volume} {11}},\
  \bibinfo {pages} {023023} (\bibinfo {year} {2009})}\BibitemShut {NoStop}%
\bibitem [{\citenamefont {Nanba}\ \emph {et~al.}(1993)\citenamefont {Nanba},
  \citenamefont {Tomikawa}, \citenamefont {Mori}, \citenamefont {Shino},
  \citenamefont {Imada}, \citenamefont {Suga}, \citenamefont {Kimura},\ and\
  \citenamefont {Kunii}}]{YbB6_nanba_1993}%
  \BibitemOpen
  \bibfield  {author} {\bibinfo {author} {\bibfnamefont {T.}~\bibnamefont
  {Nanba}}, \bibinfo {author} {\bibfnamefont {M.}~\bibnamefont {Tomikawa}},
  \bibinfo {author} {\bibfnamefont {Y.}~\bibnamefont {Mori}}, \bibinfo {author}
  {\bibfnamefont {N.}~\bibnamefont {Shino}}, \bibinfo {author} {\bibfnamefont
  {S.}~\bibnamefont {Imada}}, \bibinfo {author} {\bibfnamefont
  {S.}~\bibnamefont {Suga}}, \bibinfo {author} {\bibfnamefont {S.}~\bibnamefont
  {Kimura}},\ and\ \bibinfo {author} {\bibfnamefont {S.}~\bibnamefont
  {Kunii}},\ }\bibfield  {title} {\bibinfo {title} {Valency of {YbB}$_6$},\
  }\href {https://doi.org/10.1016/0921-4526(93)90633-H} {\bibfield  {journal}
  {\bibinfo  {journal} {Physica B Condens. Matter.}\ }\textbf {\bibinfo
  {volume} {186-188}},\ \bibinfo {pages} {557} (\bibinfo {year}
  {1993})}\BibitemShut {NoStop}%
\bibitem [{\citenamefont {Gavilano}\ \emph {et~al.}(2003)\citenamefont
  {Gavilano}, \citenamefont {Mushkolaj}, \citenamefont {Rau}, \citenamefont
  {Ott}, \citenamefont {Bianchi},\ and\ \citenamefont {Fisk}}]{Gavilano2003}%
  \BibitemOpen
  \bibfield  {author} {\bibinfo {author} {\bibfnamefont {J.}~\bibnamefont
  {Gavilano}}, \bibinfo {author} {\bibfnamefont {S.}~\bibnamefont {Mushkolaj}},
  \bibinfo {author} {\bibfnamefont {D.}~\bibnamefont {Rau}}, \bibinfo {author}
  {\bibfnamefont {H.}~\bibnamefont {Ott}}, \bibinfo {author} {\bibfnamefont
  {A.}~\bibnamefont {Bianchi}},\ and\ \bibinfo {author} {\bibfnamefont
  {Z.}~\bibnamefont {Fisk}},\ }\bibfield  {title} {\bibinfo {title} {{NMR
  studies of YbB$_6$}},\ }\href {https://doi.org/10.1016/S0921-4526(02)02445-6}
  {\bibfield  {journal} {\bibinfo  {journal} {Physica B: Condens. Matter.}\
  }\textbf {\bibinfo {volume} {329-333}},\ \bibinfo {pages} {570} (\bibinfo
  {year} {2003})}\BibitemShut {NoStop}%
\bibitem [{\citenamefont {Zhou}\ \emph {et~al.}(2015)\citenamefont {Zhou},
  \citenamefont {Kim}, \citenamefont {Rosa}, \citenamefont {Wu}, \citenamefont
  {Guo}, \citenamefont {Zhang}, \citenamefont {Wang}, \citenamefont {Kang},
  \citenamefont {Yi}, \citenamefont {Li}, \citenamefont {Li}, \citenamefont
  {Liu}, \citenamefont {Duan}, \citenamefont {Zi}, \citenamefont {Wei},
  \citenamefont {Jiang}, \citenamefont {Huang}, \citenamefont {Yang},
  \citenamefont {Fisk}, \citenamefont {Sun},\ and\ \citenamefont
  {Zhao}}]{Zhou2015}%
  \BibitemOpen
  \bibfield  {author} {\bibinfo {author} {\bibfnamefont {Y.}~\bibnamefont
  {Zhou}}, \bibinfo {author} {\bibfnamefont {D.-J.}\ \bibnamefont {Kim}},
  \bibinfo {author} {\bibfnamefont {P.~F.~S.}\ \bibnamefont {Rosa}}, \bibinfo
  {author} {\bibfnamefont {Q.}~\bibnamefont {Wu}}, \bibinfo {author}
  {\bibfnamefont {J.}~\bibnamefont {Guo}}, \bibinfo {author} {\bibfnamefont
  {S.}~\bibnamefont {Zhang}}, \bibinfo {author} {\bibfnamefont
  {Z.}~\bibnamefont {Wang}}, \bibinfo {author} {\bibfnamefont {D.}~\bibnamefont
  {Kang}}, \bibinfo {author} {\bibfnamefont {W.}~\bibnamefont {Yi}}, \bibinfo
  {author} {\bibfnamefont {Y.}~\bibnamefont {Li}}, \bibinfo {author}
  {\bibfnamefont {X.}~\bibnamefont {Li}}, \bibinfo {author} {\bibfnamefont
  {J.}~\bibnamefont {Liu}}, \bibinfo {author} {\bibfnamefont {P.}~\bibnamefont
  {Duan}}, \bibinfo {author} {\bibfnamefont {M.}~\bibnamefont {Zi}}, \bibinfo
  {author} {\bibfnamefont {X.}~\bibnamefont {Wei}}, \bibinfo {author}
  {\bibfnamefont {Z.}~\bibnamefont {Jiang}}, \bibinfo {author} {\bibfnamefont
  {Y.}~\bibnamefont {Huang}}, \bibinfo {author} {\bibfnamefont {Y.-f.}\
  \bibnamefont {Yang}}, \bibinfo {author} {\bibfnamefont {Z.}~\bibnamefont
  {Fisk}}, \bibinfo {author} {\bibfnamefont {L.}~\bibnamefont {Sun}},\ and\
  \bibinfo {author} {\bibfnamefont {Z.}~\bibnamefont {Zhao}},\ }\bibfield
  {title} {\bibinfo {title} {{Pressure-induced quantum phase transitions in a
  $\mathrm{Yb}{\mathrm{B}}_{6}$ single crystal}},\ }\href
  {https://doi.org/10.1103/PhysRevB.92.241118} {\bibfield  {journal} {\bibinfo
  {journal} {Phys. Rev. B}\ }\textbf {\bibinfo {volume} {92}},\ \bibinfo
  {pages} {241118} (\bibinfo {year} {2015})}\BibitemShut {NoStop}%
\bibitem [{\citenamefont {Majumdar}\ and\ \citenamefont
  {Balasubramanian}(2004)}]{nb4_majumdar_2004}%
  \BibitemOpen
  \bibfield  {author} {\bibinfo {author} {\bibfnamefont {D.}~\bibnamefont
  {Majumdar}}\ and\ \bibinfo {author} {\bibfnamefont {K.}~\bibnamefont
  {Balasubramanian}},\ }\bibfield  {title} {\bibinfo {title} {Theoretical study
  of the electronic states of {Nb}$_4$, {Nb}$_5$ clusters and their anions
  ({Nb}$^{-}_{4}$, {Nb}$^{-}_{5}$)},\ }\href
  {https://doi.org/10.1063/1.1769358} {\bibfield  {journal} {\bibinfo
  {journal} {J. Chem. Phys.}\ }\textbf {\bibinfo {volume} {121}},\ \bibinfo
  {pages} {4014} (\bibinfo {year} {2004})}\BibitemShut {NoStop}%
\bibitem [{\citenamefont {Sharma}\ \emph {et~al.}(2014)\citenamefont {Sharma},
  \citenamefont {Sivalingam}, \citenamefont {Neese},\ and\ \citenamefont
  {Chan}}]{fe4s4_sharma_14}%
  \BibitemOpen
  \bibfield  {author} {\bibinfo {author} {\bibfnamefont {S.}~\bibnamefont
  {Sharma}}, \bibinfo {author} {\bibfnamefont {K.}~\bibnamefont {Sivalingam}},
  \bibinfo {author} {\bibfnamefont {F.}~\bibnamefont {Neese}},\ and\ \bibinfo
  {author} {\bibfnamefont {G.~K.-L.}\ \bibnamefont {Chan}},\ }\bibfield
  {title} {\bibinfo {title} {Low-energy spectrum of iron–sulfur clusters
  directly from many-particle quantum mechanics},\ }\href
  {https://doi.org/10.1038/nchem.2041} {\bibfield  {journal} {\bibinfo
  {journal} {Nat. Chem.}\ }\textbf {\bibinfo {volume} {6}},\ \bibinfo {pages}
  {927} (\bibinfo {year} {2014})}\BibitemShut {NoStop}%
\bibitem [{\citenamefont {Presti}\ \emph {et~al.}(2019)\citenamefont {Presti},
  \citenamefont {Stoneburner}, \citenamefont {Truhlar},\ and\ \citenamefont
  {Gagliardi}}]{fe4s4_presti_19}%
  \BibitemOpen
  \bibfield  {author} {\bibinfo {author} {\bibfnamefont {D.}~\bibnamefont
  {Presti}}, \bibinfo {author} {\bibfnamefont {S.~J.}\ \bibnamefont
  {Stoneburner}}, \bibinfo {author} {\bibfnamefont {D.~G.}\ \bibnamefont
  {Truhlar}},\ and\ \bibinfo {author} {\bibfnamefont {L.}~\bibnamefont
  {Gagliardi}},\ }\bibfield  {title} {\bibinfo {title} {Full correlation in a
  multiconfigurational study of bimetallic clusters: Restricted active space
  pair-density functional theory study of [2{Fe}--2{S}] systems},\ }\href
  {https://doi.org/10.1021/acs.jpcc.9b00222} {\bibfield  {journal} {\bibinfo
  {journal} {J. Phys. Chem. C}\ }\textbf {\bibinfo {volume} {123}},\ \bibinfo
  {pages} {11899} (\bibinfo {year} {2019})}\BibitemShut {NoStop}%
\bibitem [{\citenamefont {Hozoi}\ \emph {et~al.}(2020)\citenamefont {Hozoi},
  \citenamefont {Eldeeb},\ and\ \citenamefont
  {R{\"o}{\ss}ler}}]{V148_hozoi_20}%
  \BibitemOpen
  \bibfield  {author} {\bibinfo {author} {\bibfnamefont {L.}~\bibnamefont
  {Hozoi}}, \bibinfo {author} {\bibfnamefont {M.~S.}\ \bibnamefont {Eldeeb}},\
  and\ \bibinfo {author} {\bibfnamefont {U.~K.}\ \bibnamefont
  {R{\"o}{\ss}ler}},\ }\bibfield  {title} {\bibinfo {title} {{V}$_4$
  tetrahedral units in {AV}$_4${X}$_8$ lacunar spinels: Near degeneracy, charge
  fluctuations, and configurational mixing within a valence space of up to 21
  $d$ orbitals},\ }\href {https://doi.org/10.1103/PhysRevResearch.2.022017}
  {\bibfield  {journal} {\bibinfo  {journal} {Phys. Rev. Res.}\ }\textbf
  {\bibinfo {volume} {2}},\ \bibinfo {pages} {022017} (\bibinfo {year}
  {2020})}\BibitemShut {NoStop}%
\bibitem [{\citenamefont {Petersen}\ \emph {et~al.}(2022)\citenamefont
  {Petersen}, \citenamefont {Prodan}, \citenamefont {Tsurkan}, \citenamefont
  {Krug~von Nidda}, \citenamefont {Kézsmárki}, \citenamefont {Rößler},\
  and\ \citenamefont {Hozoi}}]{Petersen2022}%
  \BibitemOpen
  \bibfield  {author} {\bibinfo {author} {\bibfnamefont {T.}~\bibnamefont
  {Petersen}}, \bibinfo {author} {\bibfnamefont {L.}~\bibnamefont {Prodan}},
  \bibinfo {author} {\bibfnamefont {V.}~\bibnamefont {Tsurkan}}, \bibinfo
  {author} {\bibfnamefont {H.-A.}\ \bibnamefont {Krug~von Nidda}}, \bibinfo
  {author} {\bibfnamefont {I.}~\bibnamefont {Kézsmárki}}, \bibinfo {author}
  {\bibfnamefont {U.~K.}\ \bibnamefont {Rößler}},\ and\ \bibinfo {author}
  {\bibfnamefont {L.}~\bibnamefont {Hozoi}},\ }\bibfield  {title} {\bibinfo
  {title} {{How Correlations and Spin–Orbit Coupling Work within Extended
  Orbitals of Transition-Metal Tetrahedra of 4d/5d Lacunar Spinels}},\ }\href
  {https://doi.org/10.1021/acs.jpclett.1c04100} {\bibfield  {journal} {\bibinfo
   {journal} {J. Phys. Chem. Lett.}\ }\textbf {\bibinfo {volume} {13}},\
  \bibinfo {pages} {1681} (\bibinfo {year} {2022})}\BibitemShut {NoStop}%
\bibitem [{\citenamefont {Fulde}(2019)}]{wfs_fulde_2019}%
  \BibitemOpen
  \bibfield  {author} {\bibinfo {author} {\bibfnamefont {P.}~\bibnamefont
  {Fulde}},\ }\bibfield  {title} {\bibinfo {title} {Wavefunctions of
  macroscopic electron systems},\ }\href {https://doi.org/10.1063/1.5050329}
  {\bibfield  {journal} {\bibinfo  {journal} {J. Chem. Phys.}\ }\textbf
  {\bibinfo {volume} {150}},\ \bibinfo {pages} {030901} (\bibinfo {year}
  {2019})}\BibitemShut {NoStop}%
\bibitem [{\citenamefont {Derenzo}\ \emph {et~al.}(2000)\citenamefont
  {Derenzo}, \citenamefont {Klintenberg},\ and\ \citenamefont
  {Weber}}]{Derenzo2000}%
  \BibitemOpen
  \bibfield  {author} {\bibinfo {author} {\bibfnamefont {S.~E.}\ \bibnamefont
  {Derenzo}}, \bibinfo {author} {\bibfnamefont {M.~K.}\ \bibnamefont
  {Klintenberg}},\ and\ \bibinfo {author} {\bibfnamefont {M.~J.}\ \bibnamefont
  {Weber}},\ }\bibfield  {title} {\bibinfo {title} {Determining point charge
  arrays that produce accurate ionic crystal fields for atomic cluster
  calculations},\ }\href {https://doi.org/10.1063/1.480776} {\bibfield
  {journal} {\bibinfo  {journal} {J. Chem. Phys.}\ }\textbf {\bibinfo {volume}
  {112}},\ \bibinfo {pages} {2074} (\bibinfo {year} {2000})}\BibitemShut
  {NoStop}%
\bibitem [{\citenamefont {Klintenberg}\ \emph {et~al.}(2000)\citenamefont
  {Klintenberg}, \citenamefont {Derenzo},\ and\ \citenamefont
  {Weber}}]{Klintenberg2000}%
  \BibitemOpen
  \bibfield  {author} {\bibinfo {author} {\bibfnamefont {M.}~\bibnamefont
  {Klintenberg}}, \bibinfo {author} {\bibfnamefont {S.}~\bibnamefont
  {Derenzo}},\ and\ \bibinfo {author} {\bibfnamefont {M.}~\bibnamefont
  {Weber}},\ }\bibfield  {title} {\bibinfo {title} {Accurate crystal fields for
  embedded cluster calculations},\ }\href
  {https://doi.org/https://doi.org/10.1016/S0010-4655(00)00071-0} {\bibfield
  {journal} {\bibinfo  {journal} {Comp. Phys. Commun.}\ }\textbf {\bibinfo
  {volume} {131}},\ \bibinfo {pages} {120} (\bibinfo {year}
  {2000})}\BibitemShut {NoStop}%
\bibitem [{\citenamefont {Schmitt}\ \emph {et~al.}(2001)\citenamefont
  {Schmitt}, \citenamefont {Stückl}, \citenamefont {Ripplinger},\ and\
  \citenamefont {Albert}}]{Schmitt2001}%
  \BibitemOpen
  \bibfield  {author} {\bibinfo {author} {\bibfnamefont {K.}~\bibnamefont
  {Schmitt}}, \bibinfo {author} {\bibfnamefont {C.}~\bibnamefont {Stückl}},
  \bibinfo {author} {\bibfnamefont {H.}~\bibnamefont {Ripplinger}},\ and\
  \bibinfo {author} {\bibfnamefont {B.}~\bibnamefont {Albert}},\ }\bibfield
  {title} {\bibinfo {title} {Crystal and electronic structure of {BaB$_6$} in
  comparison with {CaB$_6$} and molecular [{B$_6$H$_6$}]$^{2-}$},\ }\href
  {https://doi.org/10.1016/S1293-2558(00)01091-8} {\bibfield  {journal}
  {\bibinfo  {journal} {Solid State Sci.}\ }\textbf {\bibinfo {volume} {3}},\
  \bibinfo {pages} {321} (\bibinfo {year} {2001})}\BibitemShut {NoStop}%
\bibitem [{\citenamefont {Lee}\ and\ \citenamefont {Wang}(2005)}]{Lee2005}%
  \BibitemOpen
  \bibfield  {author} {\bibinfo {author} {\bibfnamefont {B.}~\bibnamefont
  {Lee}}\ and\ \bibinfo {author} {\bibfnamefont {L.-W.}\ \bibnamefont {Wang}},\
  }\bibfield  {title} {\bibinfo {title} {Electronic structure of calcium
  hexaborides},\ }\href {https://doi.org/10.1063/1.2150578} {\bibfield
  {journal} {\bibinfo  {journal} {Appl. Phys. Lett.}\ }\textbf {\bibinfo
  {volume} {87}},\ \bibinfo {pages} {262509} (\bibinfo {year}
  {2005})}\BibitemShut {NoStop}%
\bibitem [{\citenamefont {Schmidt}\ \emph {et~al.}(2015)\citenamefont
  {Schmidt}, \citenamefont {Buettner}, \citenamefont {Graeve},\ and\
  \citenamefont {Vasquez}}]{Schmidt2015}%
  \BibitemOpen
  \bibfield  {author} {\bibinfo {author} {\bibfnamefont {K.~M.}\ \bibnamefont
  {Schmidt}}, \bibinfo {author} {\bibfnamefont {A.~B.}\ \bibnamefont
  {Buettner}}, \bibinfo {author} {\bibfnamefont {O.~A.}\ \bibnamefont
  {Graeve}},\ and\ \bibinfo {author} {\bibfnamefont {V.~R.}\ \bibnamefont
  {Vasquez}},\ }\bibfield  {title} {\bibinfo {title} {Interatomic pair
  potentials from {DFT} and molecular dynamics for {Ca}, {Ba}, and {Sr}
  hexaborides},\ }\href {https://doi.org/10.1039/C5TC01398D} {\bibfield
  {journal} {\bibinfo  {journal} {J. Mater. Chem. C}\ }\textbf {\bibinfo
  {volume} {3}},\ \bibinfo {pages} {8649} (\bibinfo {year} {2015})}\BibitemShut
  {NoStop}%
\bibitem [{\citenamefont {Jun}\ \emph {et~al.}(2007)\citenamefont {Jun},
  \citenamefont {Jiang},\ and\ \citenamefont {Lemin}}]{Jun2007}%
  \BibitemOpen
  \bibfield  {author} {\bibinfo {author} {\bibfnamefont {J.}~\bibnamefont
  {Jun}}, \bibinfo {author} {\bibfnamefont {B.}~\bibnamefont {Jiang}},\ and\
  \bibinfo {author} {\bibfnamefont {L.}~\bibnamefont {Lemin}},\ }\bibfield
  {title} {\bibinfo {title} {Study on band structure of {YbB$_6$} and analysis
  of its optical conductivity spectrum},\ }\href
  {https://doi.org/10.1016/S1002-0721(08)60002-2} {\bibfield  {journal}
  {\bibinfo  {journal} {J. Rare Earth}\ }\textbf {\bibinfo {volume} {25}},\
  \bibinfo {pages} {654} (\bibinfo {year} {2007})}\BibitemShut {NoStop}%
\bibitem [{\citenamefont {Kaupp}\ \emph {et~al.}(1991)\citenamefont {Kaupp},
  \citenamefont {Schleyer}, \citenamefont {Stoll},\ and\ \citenamefont
  {Preuss}}]{Kaupp1991}%
  \BibitemOpen
  \bibfield  {author} {\bibinfo {author} {\bibfnamefont {M.}~\bibnamefont
  {Kaupp}}, \bibinfo {author} {\bibfnamefont {P.~v.~R.}\ \bibnamefont
  {Schleyer}}, \bibinfo {author} {\bibfnamefont {H.}~\bibnamefont {Stoll}},\
  and\ \bibinfo {author} {\bibfnamefont {H.}~\bibnamefont {Preuss}},\
  }\bibfield  {title} {\bibinfo {title} {Pseudopotential approaches to {Ca},
  {Sr}, and {Ba} hydrides. {Why} are some alkaline earth {MX$_2$} compounds
  bent?},\ }\href {https://doi.org/10.1063/1.459993} {\bibfield  {journal}
  {\bibinfo  {journal} {J. Chem. Phys.}\ }\textbf {\bibinfo {volume} {94}},\
  \bibinfo {pages} {1360} (\bibinfo {year} {1991})}\BibitemShut {NoStop}%
\bibitem [{\citenamefont {Dolg}\ \emph {et~al.}(1989)\citenamefont {Dolg},
  \citenamefont {Stoll},\ and\ \citenamefont {Preuss}}]{Dolg1989}%
  \BibitemOpen
  \bibfield  {author} {\bibinfo {author} {\bibfnamefont {M.}~\bibnamefont
  {Dolg}}, \bibinfo {author} {\bibfnamefont {H.}~\bibnamefont {Stoll}},\ and\
  \bibinfo {author} {\bibfnamefont {H.}~\bibnamefont {Preuss}},\ }\bibfield
  {title} {\bibinfo {title} {Energy‐adjusted ab initio pseudopotentials for
  the rare earth elements},\ }\href {https://doi.org/10.1063/1.456066}
  {\bibfield  {journal} {\bibinfo  {journal} {J. Chem. Phys.}\ }\textbf
  {\bibinfo {volume} {90}},\ \bibinfo {pages} {1730} (\bibinfo {year}
  {1989})}\BibitemShut {NoStop}%
\bibitem [{\citenamefont {Bergner}\ \emph {et~al.}(1993)\citenamefont
  {Bergner}, \citenamefont {Dolg}, \citenamefont {K{\"u}chle}, \citenamefont
  {Stoll},\ and\ \citenamefont {Preu{\ss}}}]{Bergner1993}%
  \BibitemOpen
  \bibfield  {author} {\bibinfo {author} {\bibfnamefont {A.}~\bibnamefont
  {Bergner}}, \bibinfo {author} {\bibfnamefont {M.}~\bibnamefont {Dolg}},
  \bibinfo {author} {\bibfnamefont {W.}~\bibnamefont {K{\"u}chle}}, \bibinfo
  {author} {\bibfnamefont {H.}~\bibnamefont {Stoll}},\ and\ \bibinfo {author}
  {\bibfnamefont {H.}~\bibnamefont {Preu{\ss}}},\ }\bibfield  {title} {\bibinfo
  {title} {Ab initio energy-adjusted pseudopotentials for elements of groups
  13–17},\ }\href {https://doi.org/10.1080/00268979300103121} {\bibfield
  {journal} {\bibinfo  {journal} {Mol. Phys.}\ }\textbf {\bibinfo {volume}
  {80}},\ \bibinfo {pages} {1431} (\bibinfo {year} {1993})}\BibitemShut
  {NoStop}%
\bibitem [{\citenamefont {Dunning}(1989)}]{Dunning1989}%
  \BibitemOpen
  \bibfield  {author} {\bibinfo {author} {\bibfnamefont {T.~H.}\ \bibnamefont
  {Dunning}},\ }\bibfield  {title} {\bibinfo {title} {Gaussian basis sets for
  use in correlated molecular calculations. {I.} {The} atoms boron through neon
  and hydrogen},\ }\href {https://doi.org/10.1063/1.456153} {\bibfield
  {journal} {\bibinfo  {journal} {J. Chem. Phys.}\ }\textbf {\bibinfo {volume}
  {90}},\ \bibinfo {pages} {1007} (\bibinfo {year} {1989})}\BibitemShut
  {NoStop}%
\end{thebibliography}%

\end{document}